\documentclass[nopacs,twocolumn,aps,pre,superscriptaddress,
letterpaper]{revtex4-1}
\usepackage{graphicx,bm}
\usepackage{amsmath,amssymb}

\begin{document}


\title{
Magnetically induced elastic deformations of magnetic gels and elastomers\\ containing particles of mixed size
}

\author{Lukas Fischer}
\email{lfischer@thphy.uni-duesseldorf.de}
\affiliation{Institut f{\"u}r Theoretische Physik II: Weiche Materie, 
Heinrich-Heine-Universit{\"a}t D{\"u}sseldorf, Universit\"atsstra{\ss}e 1,  D-40225 D{\"u}sseldorf, Germany}
\author{Andreas M. Menzel}
\email{menzel@thphy.uni-duesseldorf.de}
\affiliation{Institut f{\"u}r Theoretische Physik II: Weiche Materie, 
Heinrich-Heine-Universit{\"a}t D{\"u}sseldorf, Universit\"atsstra{\ss}e 1, D-40225 D{\"u}sseldorf, Germany}

\date{\today}

\begin{abstract}
Soft elastic composite materials can serve as actuators when they transform changes in external fields into mechanical deformation. Here, we address the corresponding deformational behavior of magnetic gels and elastomers, consisting of magnetizable colloidal particles in a soft polymeric matrix and exposed to external magnetic fields. Since many practical realizations of such materials involve particulate inclusions of polydisperse size distributions, we concentrate on the effect that mixed particle sizes have on the overall deformational response. To perform a systematic study, our focus is on binary size distributions. We systematically vary the fraction of larger particles relative to smaller ones and characterize the resulting magnetostrictive behavior. The consequences for systems of various different spatial particle arrangements and different degrees of compressibility of the elastic matrix are evaluated. In parts, we observe a qualitative change in the overall response for selected systems of mixed particle sizes. Specifically, overall changes in volume and relative elongations or contractions in response to an induced magnetization can be reversed into the opposite types of behavior. Our results should apply to the characteristics of other soft elastic composite materials like electrorheological gels and elastomers when exposed to external electric fields as well. Overall, we hope to stimulate the further investigation on the purposeful use of mixed particle sizes as a means to design tailored requested material behavior. 
\end{abstract}

\maketitle

\section{Introduction}

Magnetic gels and elastomers consist of magnetic or magnetizable colloidal particles locked into a soft, elastic, permanently crosslinked polymeric body \cite{jolly1996magnetoviscoelastic, filipcsei2007magnetic, ilg2013stimuli, 
li2014state, odenbach2016microstructure, menzel2015tuned, schmauch2017chained, weeber2018polymer, weeber2019studying, stolbov2019magnetostriction, menzel2019mesoscopic, schumann2019microscopic, zhou2019magnetoresponsive}. Representing a class of stimuli-responsive materials, at least two types of reaction to external magnetic fields are standing out. First, the overall mechanical properties and stiffness are affected by sufficiently strong external magnetic fields, a scenario that was termed magnetorheological effect \cite{jolly1996magnetoviscoelastic, jolly1996model, jarkova2003hydrodynamics, filipcsei2007magnetic, stepanov2007effect, bose2009magnetorheological, chertovich2010new, wood2011modeling, ivaneyko2012effects, evans2012highly, han2013field, borin2013tuning, chiba2013wide, pessot2014structural, menzel2015tuned, sorokin2015hysteresis, pessot2016dynamic, volkova2017motion, oguro2017magnetic, pessot2018tunable, watanabe2018effect}. Second, the materials tend to respond by significant elastic deformations, which allows for the construction of soft actuators and is often referred to as magnetostrictive behavior \cite{zrinyi1996deformation, filipcsei2007magnetic, gollwitzer2008measuring, fuhrer2009crosslinking, ivaneyko2011magneto, stolbov2011modelling, 
maas2016experimental, metsch2016numerical, attaran2017modeling, fischer2019magnetostriction, fischer2020towards}, particularly when the external magnetic fields are homogeneous. We here concentrate on the latter effect. 

It is already known from the study of magnetic fluids, consisting of magnetic or magnetizable colloidal particles suspended in a carrier liquid \cite{rosensweig1985ferrohydrodynamics, huke2004magnetic, odenbach2004recent, holm2005structure, vicente2011magnetorheological}, that the particle size is a key parameter. For example, it has been demonstrated that the magnetoviscous effect, that is the change in the macroscopic fluid viscosity induced by external magnetic fields, is dominated mainly by the response of the larger suspended particles \cite{thurm2003particle}. For magnetic gels and elastomers, the dependence of the material behavior on the particle size has been analyzed as well. Changes on the type of behavior with varying particle size were partially attributed to the higher rotational mobility of smaller particles in the elastic environment \cite{kramarenko2015magnetic}. Similarly, the particle size can affect the formation of structural elements when the samples are cured under an external magnetic field \cite{borbath2012xmuct}. A stronger magnetorheological effect was observed for samples containing particles of larger size \cite{stepanov2007effect, bose2009magnetorheological, sorokin2015hysteresis, winger2019influence}. 

Actual samples are frequently based on particles of polydisperse size distribution. 
Nevertheless, theoretical approaches frequently assume a uniform particle size. Examples for exceptions are finite-element simulations \cite{kalina2016microscale} or dipole-spring models \cite{pessot2018tunable}. Moreover, hybrid models investigate the behavior of discrete large particles in an elastic environment containing a lot of magnetizable small particles by approximating the latter as a magnetizable elastic continuum. 
In contrast to that, genuine macroscopic continuum theories often consider the whole system as a continuous magnetic or magnetizable medium, therefore do not resolve any actual particle sizes explicitly, but represent the resulting effects by the values of the involved material parameters \cite{jarkova2003hydrodynamics, bohlius2004macroscopic, gebhart2019general}. 

To be able to perform a systematic study of the consequences of the presence of particles of different sizes in the system, we here concentrate on particles of binary size distribution in a mesoscopic description. It is well known that in general this reduced binary type of deviation from a uniform particle size can already have strong and qualitative effects on the overall behavior, for instance in the context of colloidal glasses.
Several previous experimental studies and associated strategies of modeling on magnetic gels and elastomers concentrate on materials of a relatively bimodal size distribution of the contained inclusions \cite{stepanov2007effect, lockette2008dynamic, stepanov2009magnetorheological, chertovich2010new, melenev2011modeling, tian2011microstructure, sorokin2015hysteresis, sorokin2017magnetorheological}. It was found, for example, that bimodal size distributions can enhance the magnetorheological effect \cite{li2010study}. 

In the present study, our focus is on the influence of mixed particle sizes on the overall magnetostrictive response of the system. For this purpose, in model systems of binary particle size distributions, we systematically increase the number of smaller particles at the cost of the number of larger particles, keeping the overall particle number constant. We evaluate how the magnetostrictive behavior, appropriately rescaled to take into account the different particle sizes, changes during these variations of the size distribution. Different discrete spatial arrangements of the particles are considered. As a benefit of our theoretical work, we are able to selectively concentrate on isolated properties related to the particle size and to study their impact on the overall behavior, excluding other aspects that may play a role in real samples as well. This helps us to understand the relative importance of specific aspects. In the present case, we concentrate on the roles of the magnitude of the magnetic moment and of the displaceability within the elastic matrix as related to the particle size. Other effects, for instance variations of the magnetization behavior with the particle size, are not taken into account. 

We continue in the following way. In section~\ref{sec_model}, we provide a brief overview on the mesoscopic model system that we use to perform our evaluations, together with our protocol of introducing and modifying the binary size distribution of the magnetizable particles. Results for various different spatial particle arrangements and compressibilities of the elastic matrix are then presented in section~\ref{sec_results}. We conclude in section~\ref{sec_concl}.

\section{Mesoscopic model systems containing discrete spatial particle arrangements of binary size distributions}
\label{sec_model}

\begin{figure}
\includegraphics[width=8.3cm]{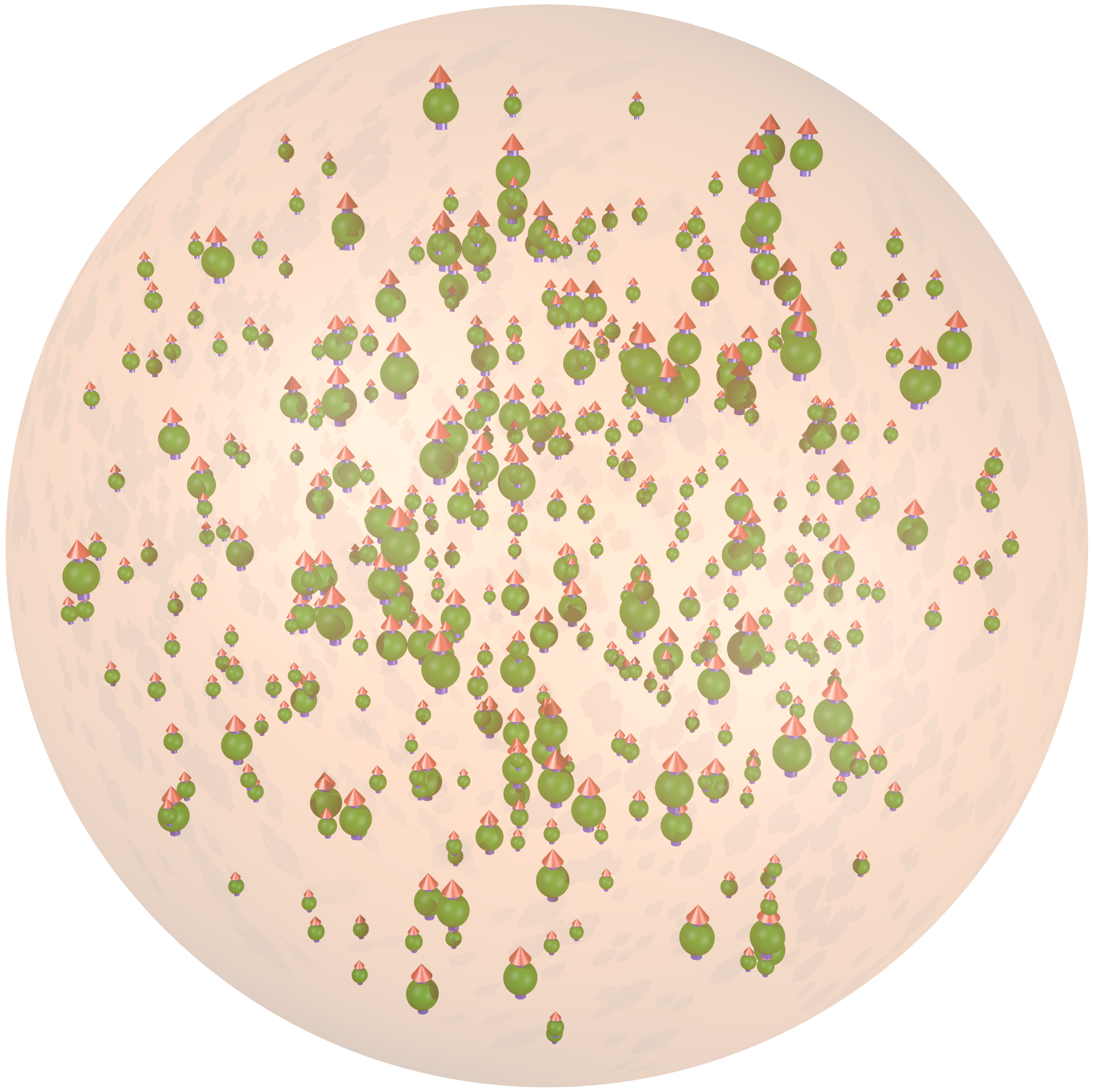}
\caption{The system considered in the present study consists of a soft spherical elastic body that contains a discrete set of magnetizable spherical particles. These particles feature a binary size distribution, implying that the diameter of the larger particles is twice the diameter of the smaller particles. Upon exposure to a strong homogeneous external magnetic field, here vertically oriented, the particles are assumed to be magnetized to saturation. We indicate the induced magnetic particle moments by the small arrows on the inclusions. In our investigation, we analyze and quantify the types of overall deformation of the enclosing elastic sphere induced by this magnetization for different spatial arrangements of the particles, different number fractions $x$ of the smaller particles, and different Poisson ratios $\nu$ of the soft elastic sphere.}
\label{fig_setup}
\end{figure}

To perform our investigations, we utilize a recently developed discrete mesoscopic model system \cite{fischer2019magnetostriction, fischer2020towards}. It allows to calculate overall mechanical deformations of a soft elastic spherical body in response to the magnetization of a discrete set of embedded spherical particles, see figure~\ref{fig_setup}. 

The radius of the elastic spherical body is denoted as $R$. Only elastic deformations of small amplitude are addressed, so that linear elasticity theory can be used for our evaluations \cite{landau1986theory}. Our analysis assumes a homogeneous isotropic elastic material forming the soft spherical body. Its elastic properties are specified by the shear modulus $\mu$ and the Poisson ratio $-1\leq\nu\leq1/2$, the latter being connected to the compressibility of the elastic matrix material \cite{landau1986theory}. 

As a major benefit of the spherical shape of the elastic matrix body, a corresponding Green's function is available to quantify its elastic deformations. This function specifies the displacements of all volume elements of the elastic body in response to a mechanical force applied at an arbitrary point within the sphere. Building on the derivation of the Green's function for an elastic sphere embedded in an infinitely extended surrounding elastic medium \cite{walpole2002elastic}, we determined this function for a free-standing elastic sphere \cite{fischer2019magnetostriction}. The explicit analytical expression is very lengthy and thus we do not reproduce it here. Due to the linearly elastic characterization, the overall response of the elastic sphere to the action of many internal force centers is obtained by simple superposition. 

In our case, it is the embedded magnetizable spherical particles that correspond to the force centers. We here include smaller particles of radius $0.01R$ and larger particles of radius $0.02R$. The number fraction of smaller particles is denoted as $x$. All particles are at least separated by a center-to-center distance of $0.11R$ from each other and by a distance of $0.06R$ of their centers from the surface of the surrounding elastic sphere. Moreover, we assume strong homogeneous saturating external magnetic fields that magnetize the systems. Thus all induced magnetic particle moments point into the same direction and only differ by their magnitudes for different particle sizes. Assuming identical material and identical internal structure of the magnetic particles, this implies an eightfold magnetic moment for the larger inclusions. 

Together, the induced magnetic interactions between the embedded particles are approximated using magnetic dipole forces. The magnetic force on particle $i$ resulting from the magnetic interaction with particle $j$ thus reads \cite{jackson1962classical}
\begin{eqnarray}
\mathbf{F}_i &=& \frac{3 \mu_0 }{4\pi {r}_{ij}^4} \Big[
\mathbf{m}_i\left(\mathbf{m}_j\cdot\mathbf{\hat{r}}_{ij}\right)
+\mathbf{m}_j\left(\mathbf{m}_i\cdot\mathbf{\hat{r}}_{ij}\right)
+\left(\mathbf{m}_i\cdot\mathbf{m}_j\right)\mathbf{\hat{r}}_{ij}
\nonumber\\
&&{}
\qquad\quad
-5\mathbf{\hat{r}}_{ij}\left(\mathbf{m}_i\cdot\mathbf{\hat{r}}_{ij}\right)\left(\mathbf{m}_j\cdot\mathbf{\hat{r}}_{ij}\right)
\Big].
\label{eq_F_magn_dipole}
\end{eqnarray}
Here, $\mathbf{m}_i$ and $\mathbf{m}_j$ are the magnetic dipole moments of particles $i$ and $j$, respectively, $\mu_0$ denotes the magnetic vacuum permeability, and $\mathbf{{r}}_{ij}
={r}_{ij}\mathbf{\hat{{r}}}_{ij}$ is the distance vector pointing from the center position of particle $j$ to the center position of particle $i$, with ${r}_{ij}=|\mathbf{{{r}}}_{ij}|$. In our implementation, we measure lengths in units of $R$ and forces in units of $\mu R^2$. 

When upon magnetization the induced magnetic forces act on the particles, the inclusions are pressed against their surrounding elastic environment and deform it \cite{puljiz2016forces, puljiz2017forces, puljiz2019displacement}. The resulting long-ranged distortions are calculated from the Green's function as mentioned above \cite{fischer2019magnetostriction, fischer2020towards}. 

As a consequence of the resulting distortions in response to the magnetic forces, the embedded magnetic particles are displaced. In turn, this couples back to the induced magnetic forces that depend on the distance vectors between the particles, see Eq.~(\ref{eq_F_magn_dipole}). This problem is solved by an iterative loop to determine the final particle positions and thus the final set of magnetic forces \cite{puljiz2016forces, fischer2019magnetostriction}. Along that way, we need to know the displacement of a single particle within the elastic sphere when a force is applied to it, as a function of the particle position and the orientation of the force. We determined corresponding fit functions as approximations for spherical particles of the two different radii \cite{fischer2019magnetostriction}. To take into account the mutual particle interactions mediated via induced distortions of the elastic body, we approximate the inclusions as point-like. This is in line with our configurations that ensure pronounced distances between the particles. 

In previous investigations, events of mutual approach of individual particles up to virtual contact under magnetic attraction were observed and analyzed \cite{stepanov2008motion, annunziata2013hardening, gundermann2014investigation, biller2014modeling, biller2015mesoscopic, gundermann2017statistical, puljiz2018reversible, goh2018dynamics}. Such a magnetomechanical collapse results when mutual magnetic attractions between individual particles surmount the elastic barrier connected to the necessary strong deformation of the elastic material between the particles. During all our investigations, we ensured that a corresponding scenario does not occur and the particles remain well separated. 

\begin{figure}
\includegraphics[width=8.3cm]{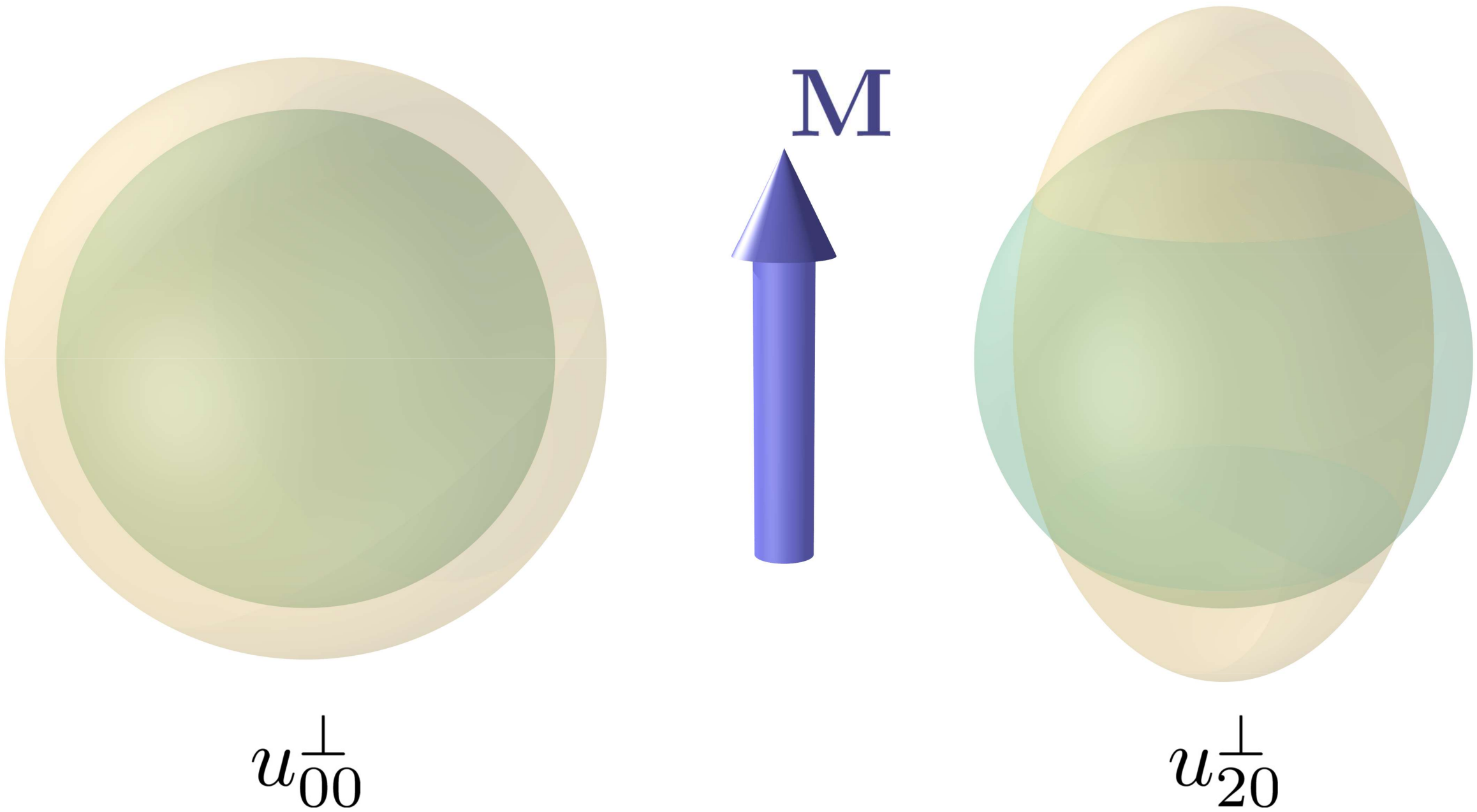}
\caption{Illustration of the types of overall deformation of the initial spherical elastic body that are quantified by the expansion coefficients $u_{00}^{\bot}$ and $u_{20}^{\bot}$. The darker (greenish) spheres indicate the initial undeformed states, while the brighter (yellowish) ellipsoids mark the deformed states. On the one hand, the coefficient $u_{00}^{\bot}$ refers to overall isotropic expansion for $u_{00}^{\bot}>0$, as indicated here, or isotropic contraction for $u_{00}^{\bot}<0$. On the other hand, $u_{20}^{\bot}>0$ represents an expansion along the axis of the magnetization $\mathbf{M}$ relative to the lateral directions, as illustrated here, while $u_{20}^{\bot}<0$ identifies a contraction along this axis relative to the lateral dimensions. For clarity, the magnitudes of deformation are indicated in an exaggerated way.}
\label{fig_expansion_coefficients}
\end{figure}

On this basis, we next determine in section~\ref{sec_results} the magnetically induced change in shape of the elastic spherical body by evaluating the resulting displacement field on the surface of the sphere \cite{fischer2019magnetostriction, fischer2020towards}. The components of the surface displacement field are expanded into spherical harmonics using the HEALPix package (http://healpix.sourceforge.net) \cite{HEALPix}. We mainly concentrate on the values of two expansion coefficients as illustrated in figure~\ref{fig_expansion_coefficients}. 

First, the expansion coefficient that we denote as $u^{\bot}_{00}$ quantifies changes in the overall volume of the spherical system. For $u^{\bot}_{00}>0$ the total volume increases, while it decreases for $u^{\bot}_{00}<0$. Second, the expansion coefficient referred to as $u^{\bot}_{20}$ quantifies changes in the overall aspect ratio. For $u^{\bot}_{20}>0$ the spherical system extends along the axis of magnetization relative to the lateral dimensions, while for $u^{\bot}_{20}<0$ a relative contraction results along this axis. A third expansion coefficient, referred to as $u^{\varphi}_{10}$, is evaluated for particle arrangements that feature an overall twist. For nonvanishing $u^{\varphi}_{10}$, a net rotation of the upper hemisphere, as selected by the magnetization direction, relative to the lower hemisphere is observed. The sign of $u^{\varphi}_{10}$ specifies the sense of this relative rotation \cite{fischer2020towards}.

\section{Results for different spatial particle arrangements and varying elastic compressibility}
\label{sec_results}

As detailed below, we now consider various different spatial arrangements of the discrete set of mesoscopic magnetizable particles embedded in the spherical elastic body. In each case, we evaluate the overall deformations as described in section~\ref{sec_model} for the number fractions $x=0$, $0.2$, $0.4$, $0.6$, $0.8$, and $1$ of the smaller particles. After fixing a specific spatial particle arrangement in the form of a regular lattice structure, there is only one possible realization for $x=0$ and $x=1$. Conversely, many different realizations are possible for the other values of $x$. The smaller and larger particles can be placed in various different ways onto the given lattice sites. Except when noted otherwise, we randomly assign the particles of different sizes to these lattice sites and average our results over $50$ realizations for each data point. In the case of the randomized structures, we additionally randomize the particle positions for each realization and again average over all $50$ systems. In this case, averages are also necessary for $x=0$ and $x=1$. 

Moreover, we evaluate our results for four different values of the Poisson ratio $\nu$ in each case. For $\nu=0.5$, the elastic body is incompressible and conserving its overall volume under any type of deformation. Thus, $u_{00}^{\bot}$ should vanish. Next, $\nu=0.3$ defines moderately compressible systems. An extreme case of compressibility is given for $\nu=0$. For this value, stretching the elastic body along one axis does not induce any lateral elastic reaction. Finally, $\nu=-0.5$ identifies a pronounced auxetic behavior, that is, the system expands to the sides when stretched along an arbitrary axis. 

The magnetic moment of the particles scales cubically with their radius. Therefore, our larger particles feature an eightfold magnetic moment when compared to the smaller particles. Thus, for identical configurations of larger particles, the resulting magnetic forces are $64$-times as strong as for the smaller particles, see Eq.~(\ref{eq_F_magn_dipole}). In our linearly elastic description, a $64$-times stronger force implies equally increased magnitudes of deformation. We therefore need to rescale our calculated quantities $u_{00}^{\bot}$, $u_{20}^{\bot}$, and $u^{\varphi}_{10}$ to make our results for different values of $x$ comparable with each other. As a divisor for rescaling, we use
\begin{eqnarray}
\lefteqn{\hspace{-.5cm}\frac{1}{N^2m^2}\sum_{i=1}^N\sum_{j=1}^N m_i m_j} 
\nonumber\\[.1cm]
&=& 
\frac{1}{N^2m^2}
\Bigg(
\sum_{i=1}^{xN}\sum_{j=1}^{xN} m_i m_j
+2\sum_{i=1}^{xN}\:\sum_{j=xN+1}^{N} m_i m_j
\nonumber\\
&&{}
\qquad\qquad+\sum_{i=xN+1}^{N}\:\sum_{j=xN+1}^{N} m_i m_j
\Bigg)
\nonumber\\[.1cm]
&=&
x^2+2x(1-x)8+(1-x)^2 8^2
\nonumber\\[.1cm]
&=&
(8-7x)^2,
\label{eq_rescale}
\end{eqnarray}
where $N$ denotes the total number of particles, $m_i=|\mathbf{m}_i|$, $m_j=|\mathbf{m}_j|$, $m$ is the magnitude of the magnetic moment of the smaller particles, while $i$ and $j$ label all particles, starting with the smaller ones. 
The rescaled quantities are denoted as $\tilde{u}_{00}^{\bot}$, $\tilde{u}_{20}^{\bot}$, and $\tilde{u}^{\varphi}_{10}$. 

Along these lines, we now analyze the deformational response upon magnetization for various different realizations of particle positioning within the spherical elastic body. Each realization contains $N\approx1000$ particles in total, unless noted otherwise. In line with realistic experimental system parameters, we set the one remaining dimensionless system parameter as $48\mu_0m^2/\pi\mu R^6=5.4\times10^{-8}$ \cite{fischer2019magnetostriction}. In all cases, we have checked that our results for $x=0$ are identical to those in our previous investigations \cite{fischer2019magnetostriction, fischer2020towards}.

\subsection{Simple cubic lattice structure}
\label{sec_sc}

\begin{figure}
\includegraphics[width=8.3cm]{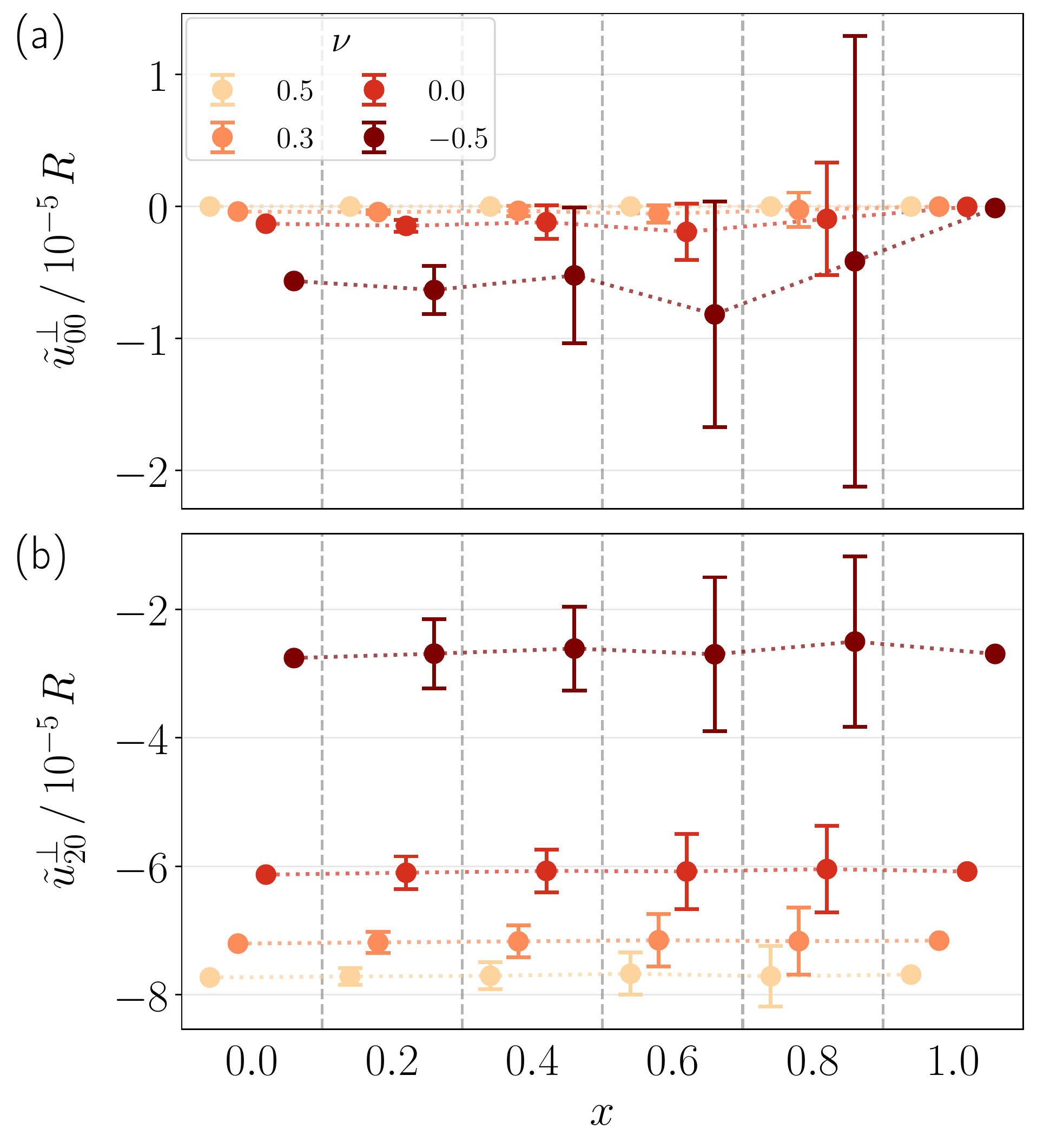}
\caption{Deformational response of soft spherical elastic bodies containing rigid particulate inclusions arranged in a simple cubic lattice-like structure. Increasing the number fraction $x$ of smaller particles implies that larger particles are randomly replaced by smaller ones. As illustrated in figure~\ref{fig_expansion_coefficients}, (a) $\tilde{u}^{\bot}_{00}$ marks changes in the overall volume, while (b) $\tilde{u}^{\bot}_{20}$ is related to elongations along the magnetization axis relative to the lateral dimensions, both quantities rescaled as given by Eq.~(\ref{eq_rescale}). Averages are taken over $50$ realizations of the systems, leading to the indicated standard deviations. For fixed $\nu<0.5$, the overall changes in volume tend to decrease in magnitude with increasing $x$. Conversely, the elongational response remains constant within the standard deviations.} 
\label{fig_sc_rand}
\end{figure}

We start by investigating systems in which the particle positioning follows regular simple cubic lattice arrangements, with magnetizations along one edge of the cubic unit cells. The fraction $x$ of smaller particles is increased by randomly replacing larger particles by smaller ones. Our results are displayed in figure~\ref{fig_sc_rand}. 

To begin, we note that, on average and for the investigated Poisson ratios $\nu<0.5$, the overall total volume upon magnetization tends to decrease, i.e., $\tilde{u}_{00}^{\bot}<0$, see figure~\ref{fig_sc_rand}(a). However, the rescaled reduction in volume decreases in magnitude with increasing $x$. Apart from that, the systems on average show a relative contraction along their axis of magnetization, as indicated by $\tilde{u}_{20}^{\bot}<0$ in figure~\ref{fig_sc_rand}(b). Interestingly, within the standard deviations, the values for $\tilde{u}_{20}^{\bot}$ remain approximately constant. First, this indicates that our way of rescaling according to Eq.~(\ref{eq_rescale}) is reasonable. Second, this result suggests that very large systems of simple cubic lattice structure may be insensitive concerning the nature of their (rescaled) response against randomized positioning of differently sized particles on their lattice points. 

Yet, we do notice the existence of the evident standard deviations in figure~\ref{fig_sc_rand}. Evidently, the auxetic systems on average are most susceptible to the presence of differently sized magnetized particles concerning resulting variations in their overall behavior. Apparently, introducing the binary size distribution can qualitatively change the response to the external magnetic field for individual realizations of the systems. For example, as indicated by the standard deviations in figure~\ref{fig_sc_rand}(a), the binary size distribution can lead to an overall expansion ($\tilde{u}_{00}^{\bot}>0$) instead of a contraction ($\tilde{u}_{00}^{\bot}<0$) of the elastic sphere for some individual realizations. This observation made us look for designed individual implementations. More precisely, for a given spatial arrangement of the particle positions, e.g., a simple cubic lattice structure, we wish to use the binary size distribution to affect the overall response by selectively replacing only larger particles on specific sites by smaller ones. 

Along these lines, we analyze the consequences of the following targeted approach. We split the set of all particle sites on the simple cubic lattice into two subsets. The site in the center of the sphere belongs to the first subset, all its nearest neighbors are part of the second subset. All the nearest neighbors of the latter again belong to the first subset, and so on. At the end, any two nearest neighbors always belong to the two different subsets. Each of these two subsets identifies octahedral structures with space diagonals along the axis of magnetization. 

\begin{figure}
\includegraphics[width=8.3cm]{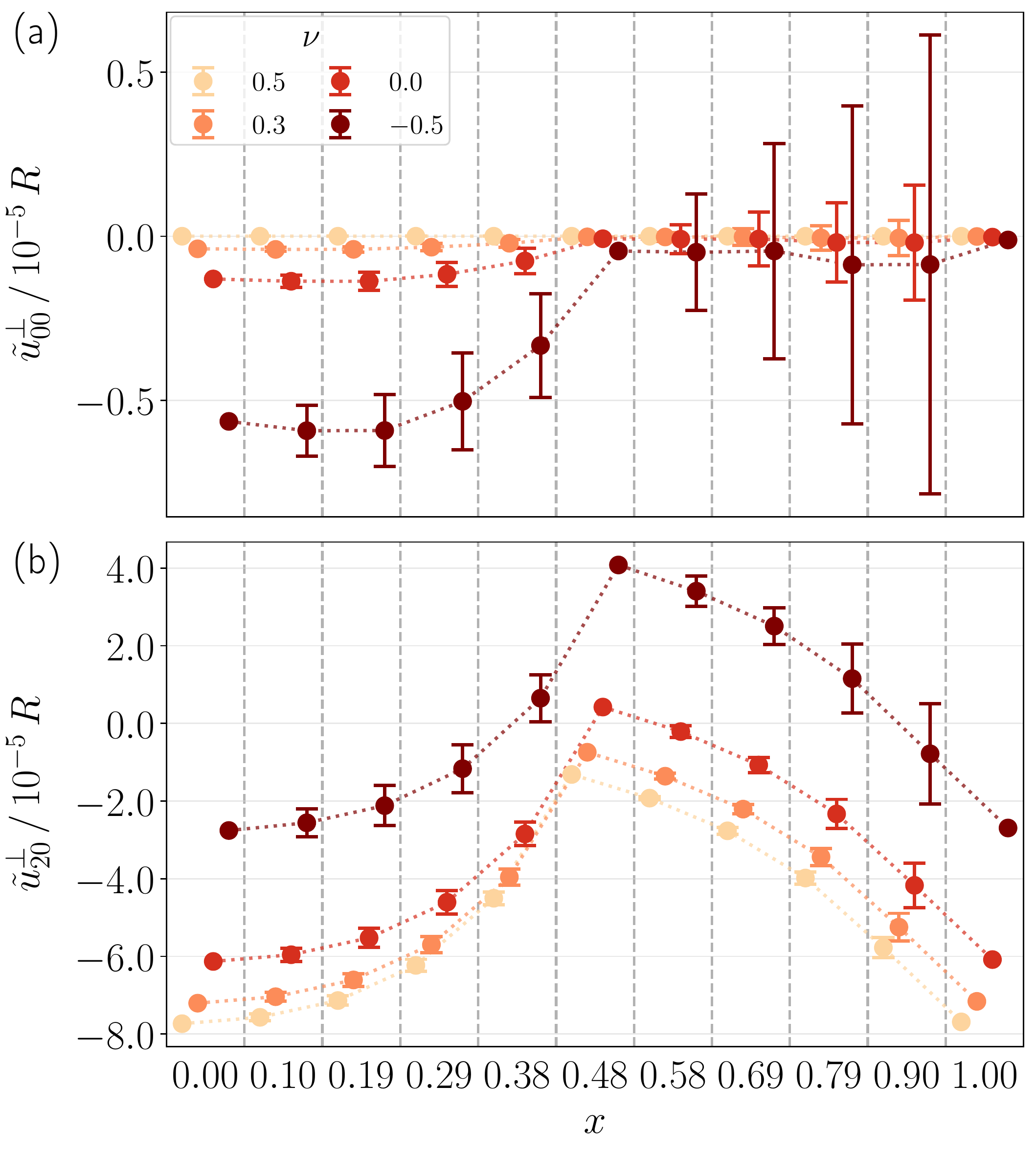}
\caption{Same as in figure~\ref{fig_sc_rand} for a simple cubic particle lattice. Yet, in contrast to figure~\ref{fig_sc_rand}, larger particles are not replaced by smaller ones in a completely random way. Instead, first only particles belonging to a specific subset are randomly exchanged, so that for $x\approx0.48$ octahedral structures of larger and smaller particles remain with space diagonals along the axis of magnetization. Subsequently, also the remaining particles are exchanged at random. Obviously, this strategy has profound consequences for the rescaled overall deformational response of the systems. For Poisson ratios $\nu=-0.5$ and $0.0$, we in between even observe relative contraction along the magnetization axis to be reversed into relative elongation.}
\label{fig_sc_target}
\end{figure}

Instead of randomly replacing any of the larger particles by a smaller one, we now first only exchange those particles at random that belong to the second subset. This has profound consequences for the rescaled deformational response upon magnetization, see figure~\ref{fig_sc_target}. 
Particularly, the behavior described by $\tilde{u}^{\bot}_{20}$ now is significantly affected by the binarization of the particle size distribution. The curves in figure~\ref{fig_sc_target}(b) first monotonically rise with increasing $x$. At the location of the maximum at $x\approx0.48$ all the larger particles of the second subset have been replaced by smaller ones. We find that the overall response of the systems under these circumstances can even be changed qualitatively. Namely, for Poisson ratios $\nu=-0.5$ and $0.0$, the overall relative contraction along the magnetization axis is reversed into relative expansion. Beyond this point, with the further increase in $x$ and now also randomly replacing particles belonging to the first subset, the curves monotonously drop. At $x=0$ and $x=1$, we find by construction the same values in figure~\ref{fig_sc_target} as in figure~\ref{fig_sc_rand}. 

This basic example already demonstrates that a tailored assignment of different particle sizes can be employed to design a requested material behavior. We continue by addressing various further types of spatial particle arrangements.

\subsection{Body-centered cubic lattice structure}
\label{sec_bcc}

\begin{figure}
\includegraphics[width=8.3cm]{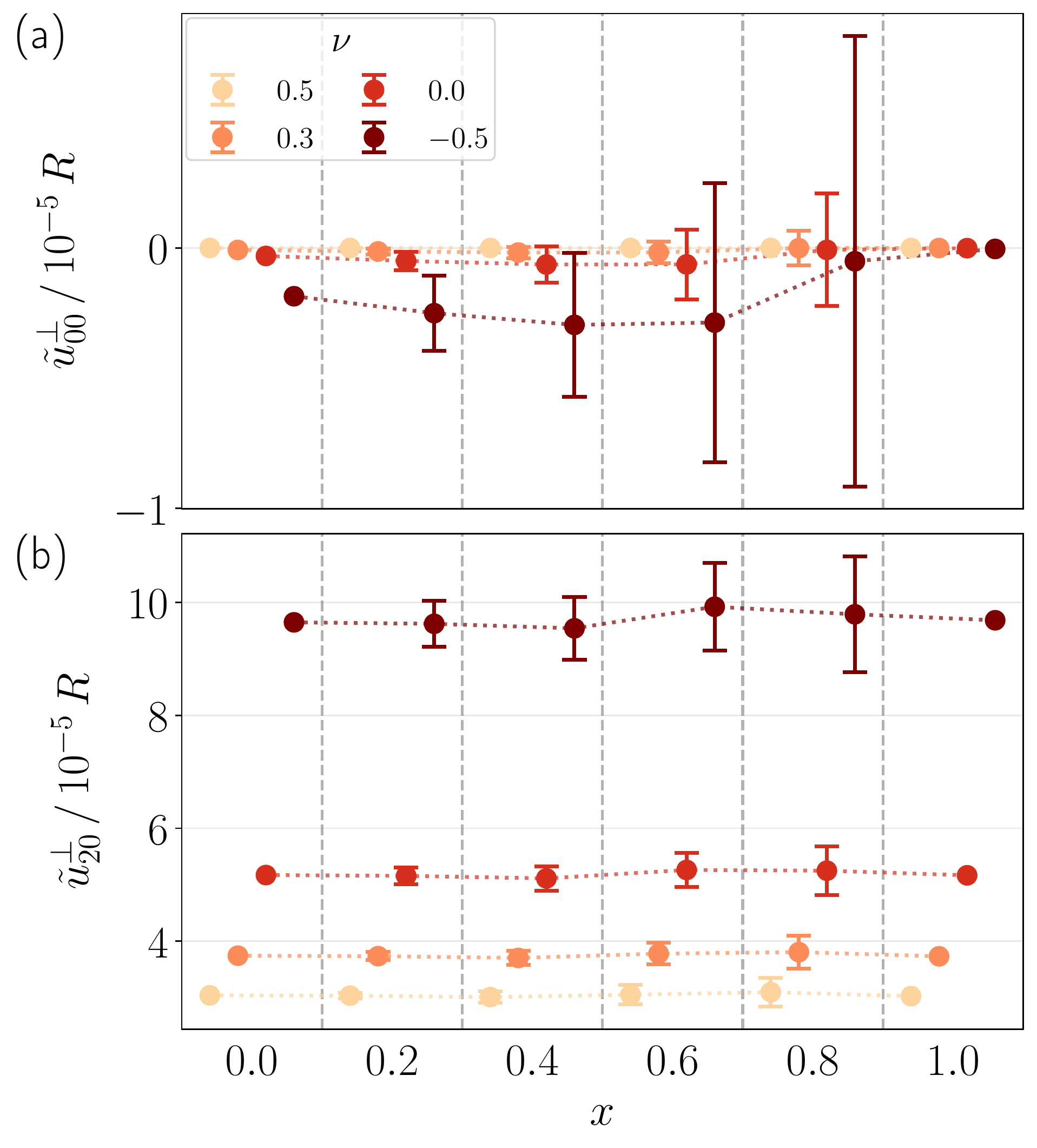}
\caption{Same as in figure~\ref{fig_sc_rand}, here for body-centered cubic particle lattices. On average, for each value of the Poisson ratio, the spherical elastic body extends along the magnetization axis. We do not find any quantitative change in this behavior as a function of $x$ within the standard deviations.}
\label{fig_bcc_rand}
\end{figure}

In contrast to simple cubic systems, body-centered cubic particle arrangements on average elongate along the magnetization axis \cite{fischer2019magnetostriction}. Again, the magnetization is directed parallel to the edge of the cubic unit cell. Apart from that, we observe for random particle replacements similar trends as for the simple cubic lattice structures, as seen by comparing figures~\ref{fig_sc_rand} and \ref{fig_bcc_rand}. 
Thus we again turn to a more specific strategy of targeted replacement of larger by smaller particles to induce a qualitative change in behavior when modifying the particle sizes.

\begin{figure}
\includegraphics[width=8.3cm]{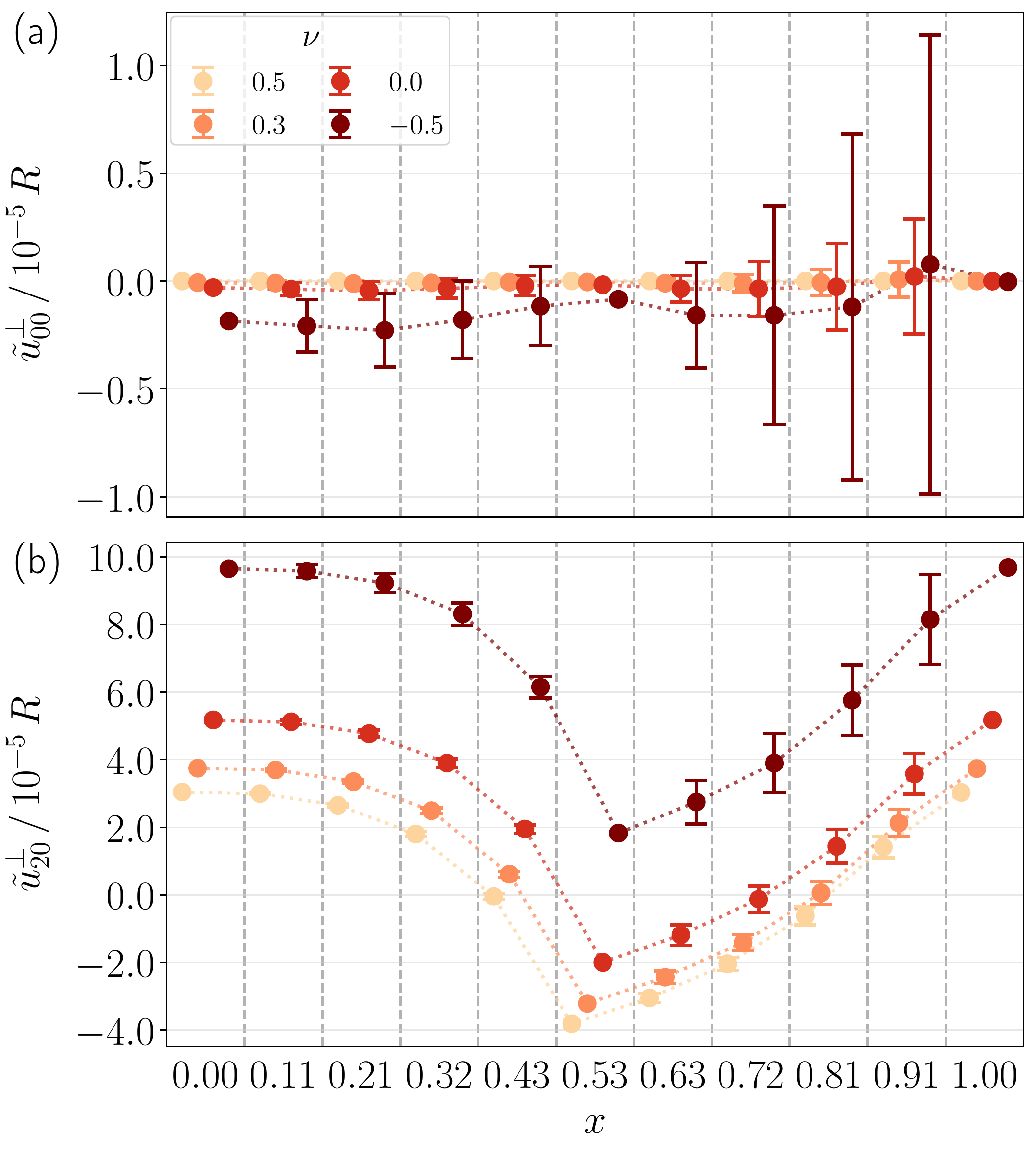}
\caption{Same as in figure~\ref{fig_bcc_rand} for body-centered cubic particle lattices, now first selectively replacing at random those larger by smaller particles that are located in the centers of the cubic unit cells. At $x\approx0.53$ all the center particles are replaced. This targeted approach implies an even qualitative change of overall response, with induced expansion along the magnetization axis being reversed into contraction for Poisson ratios $\nu=0.0$, $0.3$, and $0.5$.}
\label{fig_bcc_target}
\end{figure}

Instead of completely randomly picking larger particles that are replaced by smaller ones, we now first choose those particles at random that are located within the centers of the unit cells of our body-centered cubic structures. As figure~\ref{fig_bcc_target} demonstrates, this procedure can reverse the observed behavior. 
Specifically, the curves for $\tilde{u}^{\bot}_{20}$ now significantly drop with increasing $x$, see figure~\ref{fig_bcc_target}(b). For the Poisson ratios $\nu=0.5$, $0.3$, and $0.0$, they even decrease to negative values. This means that the magnetically induced relative extension along the axis of magnetization obtained for $x=0$ is now reversed to a relative contraction. 

When reaching $x\approx0.53$, all larger particles at the centers of the cubic unit cells have been replaced by smaller ones. Then, apparently, both the spatial arrangements of the remaining larger particles and the resulting overall responses become related to those of our simple cubic lattices studied in section~\ref{sec_sc}. Moreover, the curves in figure~\ref{fig_bcc_target}(b) at $x\approx0.53$ reach their minimum. We subsequently randomly pick the remaining larger particles for replacement, and the curves start to rise again up to $x=1$ towards values similar to those for $x=0$. By construction, at $x=0$ and $x=1$, the configurations in figure~\ref{fig_bcc_target} are identical to those at $x=0$ and $x=1$ in figure~\ref{fig_bcc_rand}, respectively.

\subsection{Face-centered cubic lattice structure}
\label{sec_fcc}

\begin{figure}
\includegraphics[width=8.3cm]{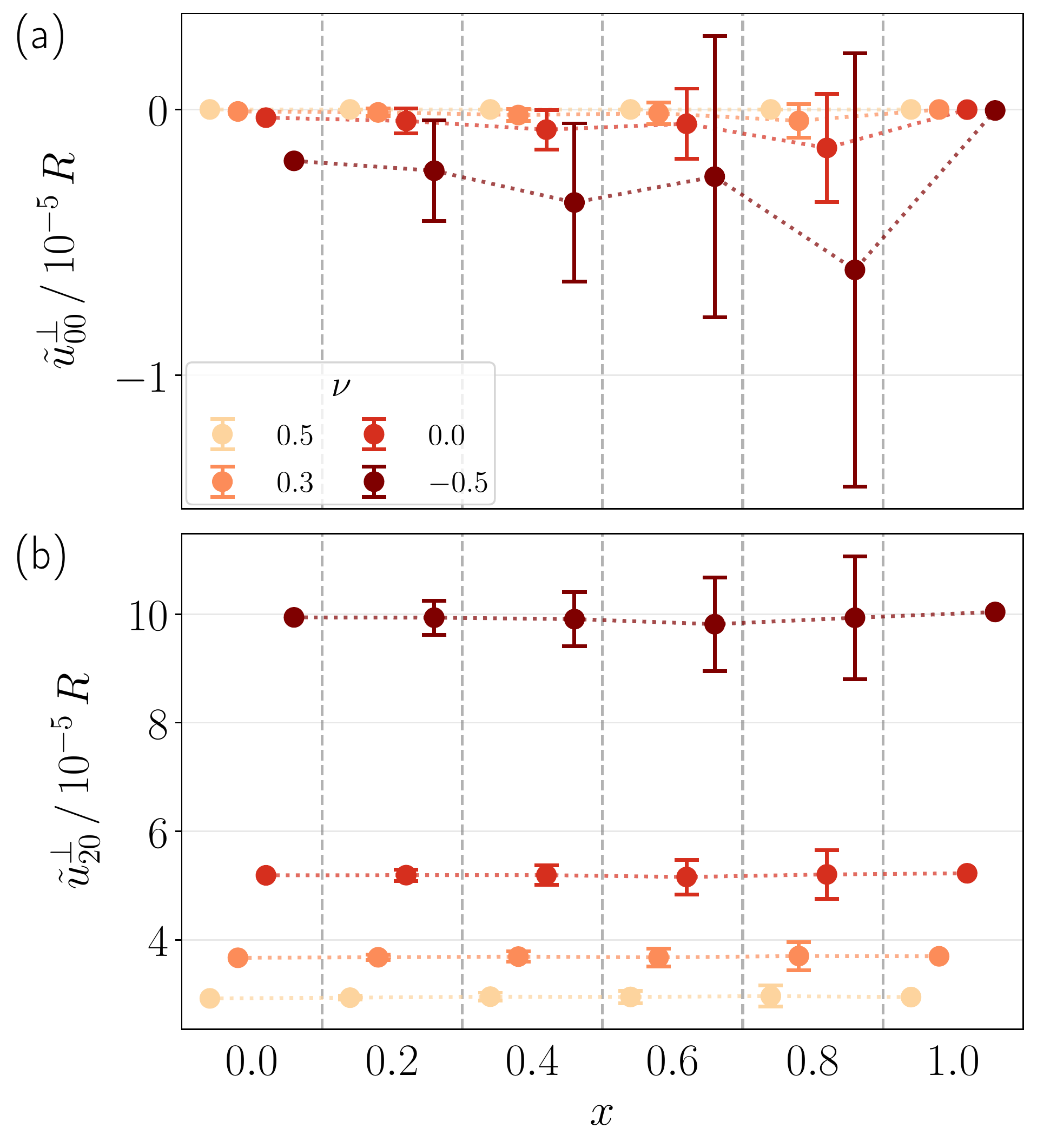}
\caption{Same as in figure~\ref{fig_sc_rand}, here for face-centered cubic particle lattices. Similar results as for the body-centered cubic particle structures in figure~\ref{fig_bcc_rand} are obtained.}
\label{fig_fcc_rand}
\end{figure}

Our results for face-centered cubic lattice structures for randomly replacing larger by smaller magnetizable particles are qualitatively similar to those for body-centered cubic lattices, as inferred by comparing figures~\ref{fig_bcc_rand} and \ref{fig_fcc_rand}. 
We therefore do not enlarge on specific observations, but directly turn to more specific results for targeted replacements of larger by smaller particles. 

\begin{figure}
\includegraphics[width=8.3cm]{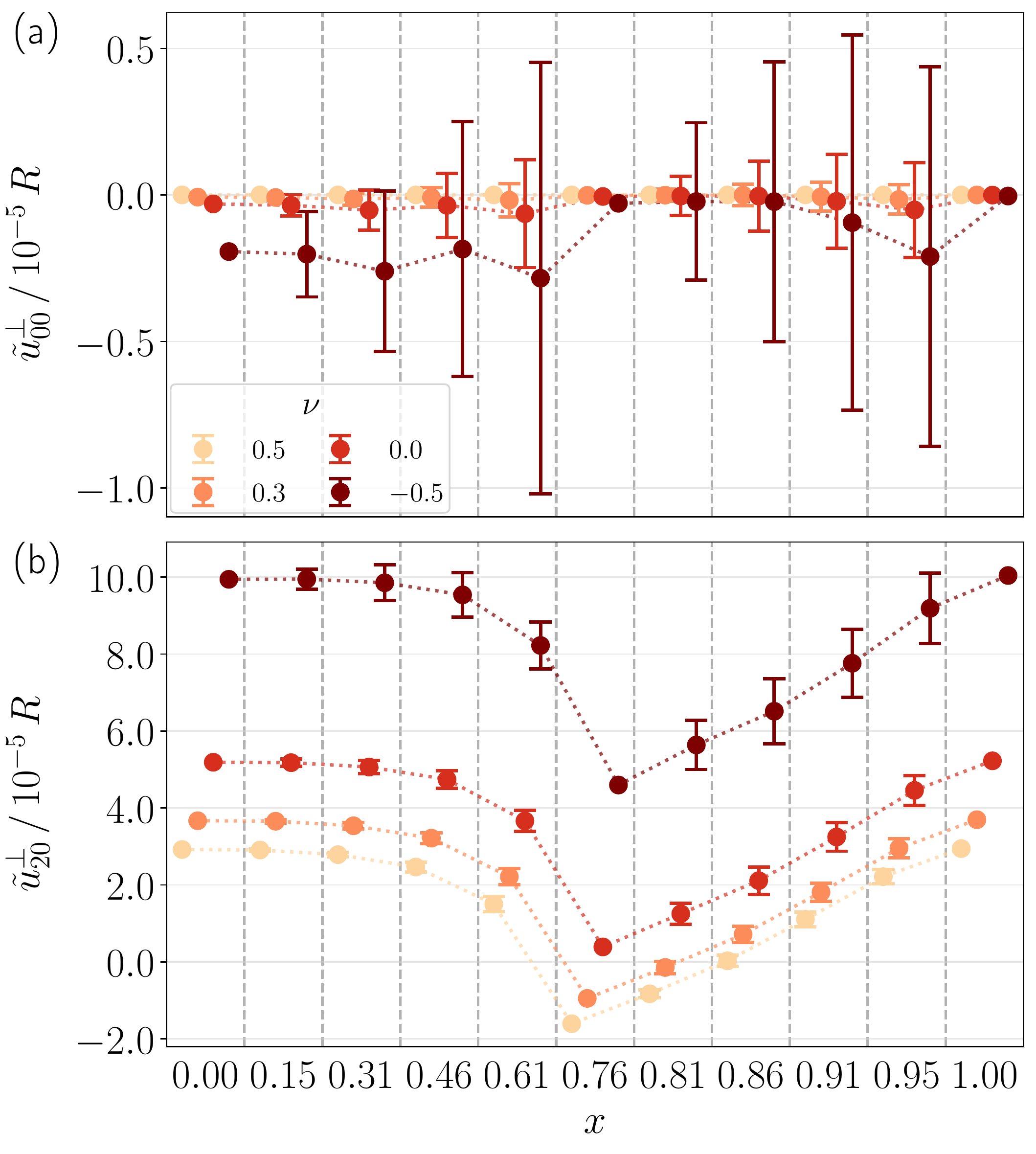}
\caption{Same as in figure~\ref{fig_fcc_rand} for face-centered cubic particle lattices, now first selectively replacing at random those larger particles by smaller ones that are located on the faces of the cubic unit cells. At $x\approx0.76$ all the particles on the faces have been exchanged. Again, an even qualitative change of the overall response can be observed following this targeted approach. In between, the induced overall expansion along the magnetization axis is reversed into contraction for Poisson ratios $\nu=0.3$ and $0.5$.}
\label{fig_fcc_target}
\end{figure}

Here, instead of completely randomly replacing larger particles by smaller ones, we first only randomly exchange those particles located on the faces of the cubic unit cells. Corresponding results are displayed in figure~\ref{fig_fcc_target}. 
As a consequence, we find a monotonous drop of the curves of $\tilde{u}_{20}^{\bot}$ in figure~\ref{fig_fcc_target}(b) with increasing $x$ up to $x\approx0.76$. At this number fraction, all larger particles on the faces of the cubic unit cells have been replaced by smaller ones. Consequently, a simple cubic lattice structure of larger particles remains. For Poisson ratios $\nu=0.3$ and $0.5$, this leads to an even qualitative change in the response. The relative elongation along the axis of magnetization is reversed into a relative contraction. Beyond the number fraction of $x\approx0.76$, the curves monotonously rise again. For $x=0$ and $x=1$, the same results are obtained as in figure~\ref{fig_fcc_rand}.

\subsection{Randomized isotropic configurations}

We now turn to basically isotropic particle distributions. In this case, the particle positions are chosen at random, only confined by the conditions listed in section~\ref{sec_model}. Out of the here-studied systems, these realizations probably correspond most closely to actual samples of well-dispersed particles that are cured in the absence of an external magnetic field. 
Depending on $x$, we randomly select a fraction of the particle positions that are assigned to the smaller instead of the larger particles. 

\begin{figure}
\includegraphics[width=8.3cm]{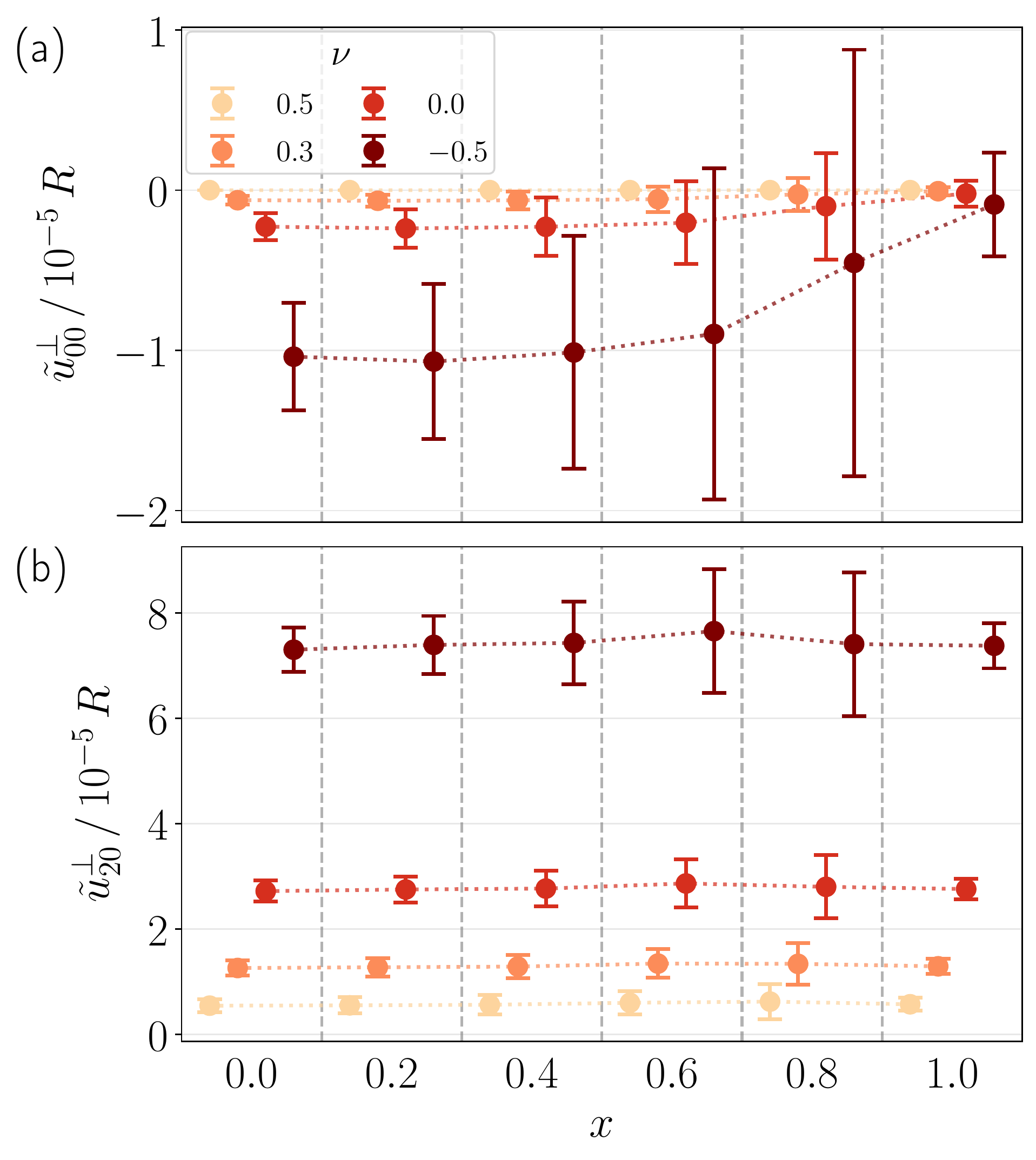}
\caption{Same as in figure~\ref{fig_sc_rand}, here for configurations of randomized particle positions. On average, for each value of the Poisson ratio, the spherical elastic body extends along the magnetization axis. Within the standard deviations, we do not observe any quantitative change in this behavior as a function of $x$.}
\label{fig_rand}
\end{figure}

We depict corresponding results in figure~\ref{fig_rand}. 
Again, we find that the rescaled change in volume, measured by $\tilde{u}_{00}^{\bot}$, tends to decrease in magnitude with increasing $x$. However, within the standard deviations, we do not observe any quantitative variation in the amount of rescaled relative elongation along the axis of magnetization indicated by $\tilde{u}_{20}^{\bot}>0$. These values agree quantitatively with our previous results obtained for uniform particle sizes in Ref.~\onlinecite{fischer2019magnetostriction}. Since the realization of randomized particle arrangements in practice most likely corresponds to fabrication methods that do not allow to control the particle positioning, we in this case also do not investigate the possibility of targeted spatial assignments of particle sizes.

\subsection{Quadratically arranged chain-like structures}
\label{sec_quadchain}

In a second step, we now turn to particle structures composed of chain-like particle aggregates. Still, within each chain, our requirements of keeping the specified minimal distances between the particles are maintained, and the initial nearest-neighbor interparticle distance within all chains is constant. Every chain is aligned parallel to the axis of magnetization. Overall, the particles still form layers, oriented normal to the magnetization axis. Since our chains are arranged according to a quadratic pattern, this implies that the particles are actually organized in a primitive tetragonal lattice. The two lattice constants perpendicular to the magnetization axis are equal, the one along the magnetization axis is smaller by a factor of $0.62$. 

\begin{figure}
\includegraphics[width=8.3cm]{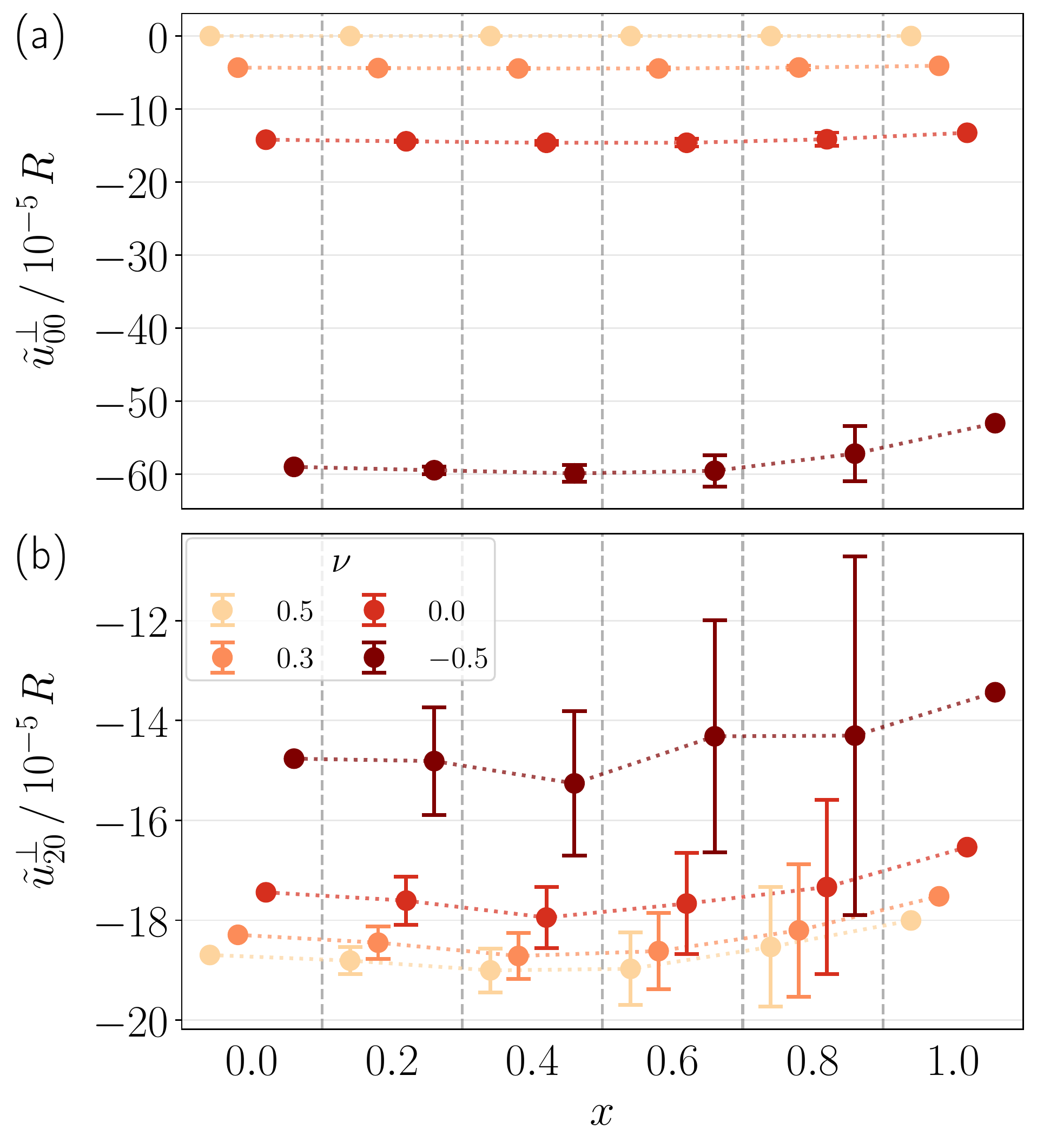}
\caption{Same as in figure~\ref{fig_sc_rand}, but here for quadratically arranged chain-like aggregates oriented along the axis of magnetization. 
For Poisson ratios $\nu<0.5$, on average, the total volume of the spherical elastic body decreases upon magnetization ($\tilde{u}_{00}^{\bot}<0$). Moreover, for each value of the Poisson ratio, the spherical elastic body on average contracts along the magnetization axis ($\tilde{u}^{\bot}_{20}<0$). The rescaled magnitude of this effect is slightly smaller for $x=1$ than for $x=0$, with weakly increased magnitudes of the averages at intermediate values of $x$, although not within the standard deviations.}
\label{fig_quadchain_rand}
\end{figure}

Upon magnetization, the spherical elastic body reduces its overall volume for the evaluated Poisson ratios $\nu<0.5$, see figure~\ref{fig_quadchain_rand}. 
Generally, this effect is not influenced as much by the random replacement of larger by smaller particles as in the previous cases, as the smaller magnitudes of the standard deviations indicate. Interesting tendencies are found for the observed rescaled relative contractions along the magnetization axis, see figure~\ref{fig_quadchain_rand}(b). First, the rescaled magnitudes of $\tilde{u}^{\bot}_{20}$ are notably smaller for $x=1$ than for $x=0$. Second, focusing on the average values of $\tilde{u}^{\bot}_{20}$, the curves tend to show a minimum at intermediate values of $x$. This implies that randomly assigning binary particle sizes can in fact increase the magnitude of the rescaled contraction along the magnetization axis. The effect is not significant within our standard deviations, yet our results imply that individual systems showing such a tendency can definitely be identified.

\subsection{Hexagonally arranged chain-like structures}
\label{sec_hex_chain}

To continue, we remain with aligned chain-like particle aggregates, now, however, arranged in a hexagonal lattice structure. In fact, this structure is related to patterns observed for real samples that are cured in the presence of a strong homogeneous external magnetic field \cite{borbath2012xmuct}. The mutual distance between the particles required in our calculation may be realized by appropriate coating of the particles or using the particles as the actual chemical crosslinkers after surface functionalization \cite{frickel2009functional, frickel2011magneto, messing2011cobalt, ilg2013stimuli}. Initially, in our calculations, within each chain-like aggregate the particles are separated from each other by a center-to-center distance of $0.11R$, while the chains themselves show a center-to-center distance of $0.19R$. 

\begin{figure}
\includegraphics[width=8.3cm]{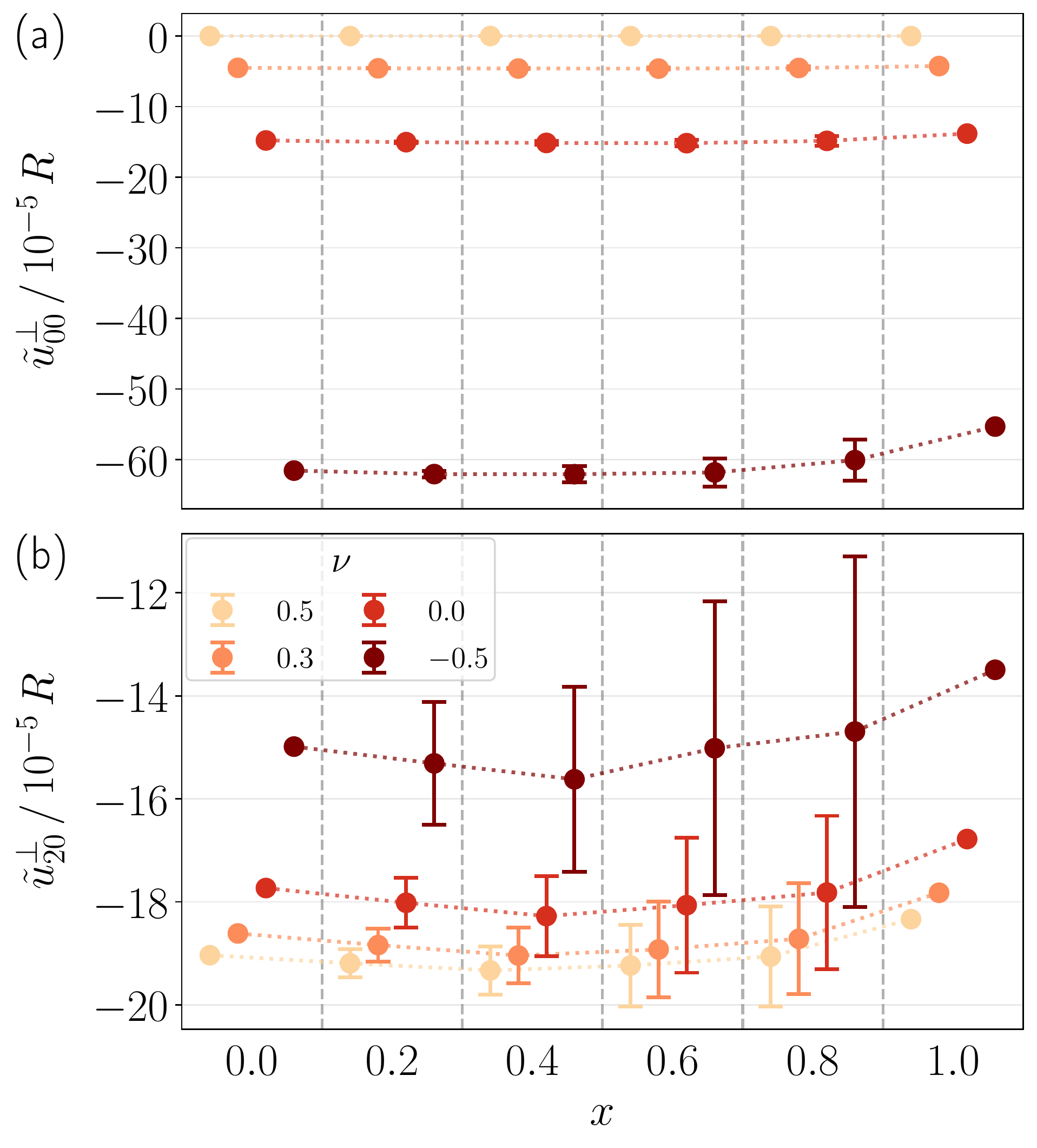}
\caption{Same as in figure~\ref{fig_quadchain_rand}, here for hexagonally arranged chain-like aggregates oriented along the axis of magnetization. Similar results as for the quadratically arranged chain-like aggregates in figure~\ref{fig_quadchain_rand} are obtained.}
\label{fig_hexchain_rand}
\end{figure}

Corresponding results for the deformation of the enclosing spherical elastic body upon magnetization along the chain-like aggregates are displayed in figure~\ref{fig_hexchain_rand}. 
They are qualitatively similar to the ones reported in section~\ref{sec_quadchain} for quadratically arranged chain-like aggregates.

\subsection{Globally twisted hexagonally arranged chain-like structures}
\label{sec_twist}

Finally, we turn to hexagonally organized chain-like structures that show an additional initial global twist. That is, an arrangement of particles similar to the one studied in section~\ref{sec_hex_chain} is initially twisted around the center axis that is parallel to the untwisted chain axes. Here, we consider a total number of only $N=623$ particles. When such a system is magnetized for not too large values of the initial twist, the structures attempt to untwist themselves. Besides the other induced types of global distortion, an overall torsional deformation results for the spherical elastic body. Therefore, corresponding systems were suggested in Ref.~\onlinecite{fischer2020towards} as candidates to realize soft torsional actuators \cite{aziz2019torsional}. 

\begin{figure}
\includegraphics[width=8.3cm]{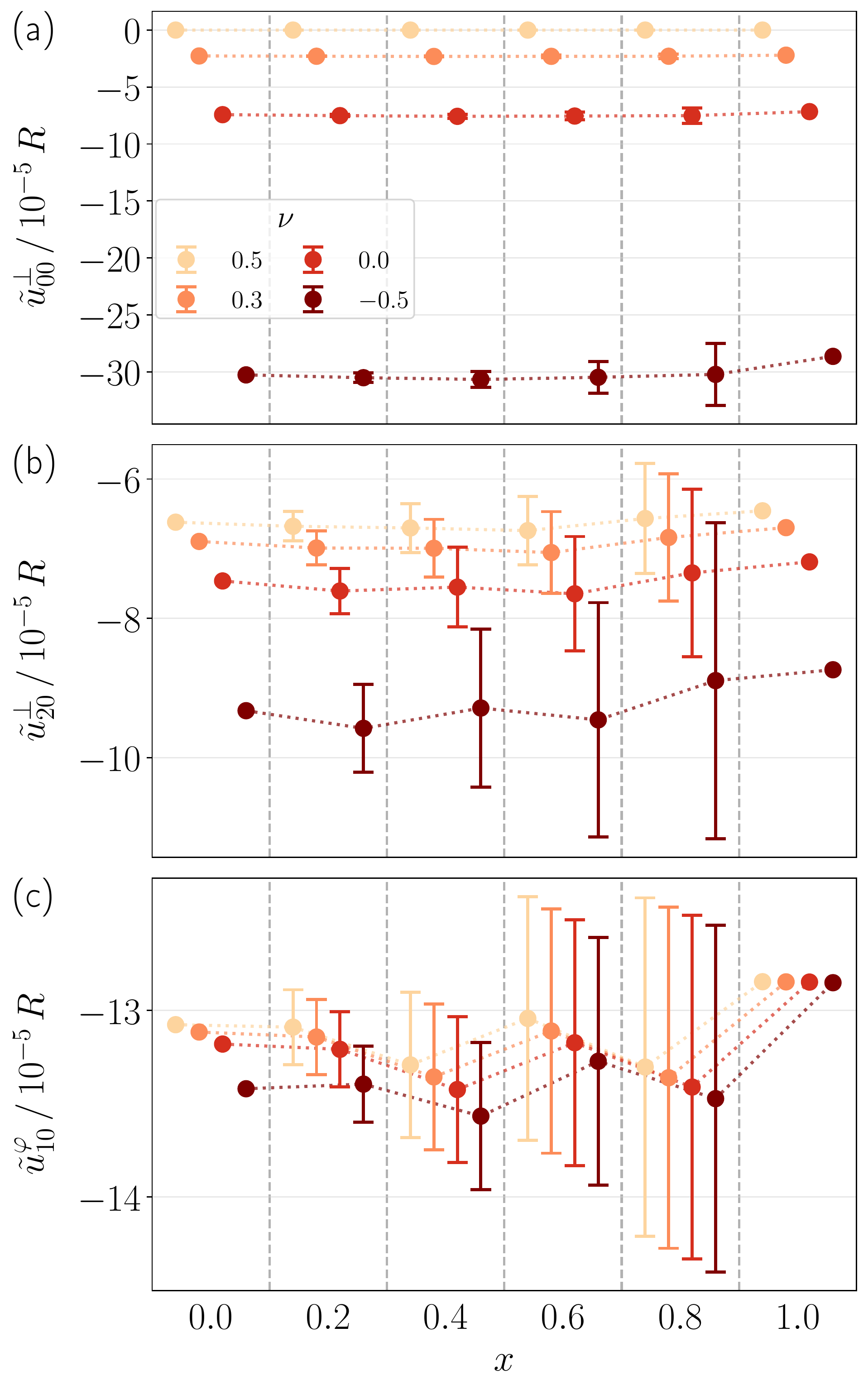}
\caption{Same as in figure~\ref{fig_hexchain_rand}, here for hexagonally arranged chain-like aggregates featuring an additional initial global twist of an approximate pitch of $11.5R$. The initial twist is implemented around the center axis that is aligned with the direction of magnetization. 
For (a) $\tilde{u}^{\bot}_{00}$ and (b) $\tilde{u}^{\bot}_{20}$ similar results as for the untwisted structures in figure~\ref{fig_hexchain_rand} are obtained, only that the magnitudes for $\tilde{u}^{\bot}_{20}$ are reversed concerning the Poisson ratios $\nu$. Additionally, (c) $\tilde{u}^{\varphi}_{10}$ quantifies the rescaled magnitude of the torsional deformation that is induced upon magnetization when the initially twisted structures attempt to untwist themselves.}
\label{fig_twist_rand}
\end{figure}

We here evaluate systems of a pitch of approximately $11.5R$, for which we observed the largest magnitude of torsional response in our previous study for a uniform particle size \cite{fischer2020towards}. As can be inferred from figures~\ref{fig_twist_rand}(a) and (b), the rescaled change in overall volume and the rescaled contraction along the axis of magnetization, respectively, behave similarly to those for the untwisted structures in figures~\ref{fig_hexchain_rand}(a) and (b) as functions of $x$. 
We note, however, that the order of the 
curves for $\tilde{u}^{\bot}_{20}$ is reversed for the investigated values of the Poisson ratio. 

In addition, we plot in figure~\ref{fig_twist_rand}(c) the coefficient $\tilde{u}^{\varphi}_{10}$, which quantifies the rescaled magnitude of the induced twist deformation. It describes the rotation of the upper hemisphere of our elastic body relative to the lower hemisphere, as seen from the direction of magnetization. The sign of $\tilde{u}^{\varphi}_{10}$ is related to the sense of this relative rotation and is thus connected to the sense of the initial twist that we impose. As figure~\ref{fig_twist_rand}(c) implies, the rescaled magnitudes of the effect are a bit smaller for the smaller particles at $x=1$ than for the larger particles at $x=0$, in line with the trends observed for $\tilde{u}^{\bot}_{00}$ and $\tilde{u}^{\bot}_{20}$ for the chain-like structures addressed in sections~\ref{sec_quadchain}--\ref{sec_twist}. 
Taking into account the magnitudes of the standard deviations, there is no clearly monotonic trend of the average values in figure~\ref{fig_twist_rand}(c) as a function of $x$.

\subsection{Layered regular structures}

From a practical point of view, it might be most realistic with presently available techniques to build up the systems containing regular particle arrangements layer by layer \cite{chen2013numerical, puljiz2016forces, puljiz2018reversible}. If these processes are automated, it may be most convenient to only use per layer one of the two particle sizes. Therefore, we add an analysis for regular particle arrangements, in which we always replace complete layers of larger particles by smaller particles in our theoretical evaluation. 

In each case, we start from the center plane normal to the magnetization direction and replace all particles within this plane. Then, from there, we additionally replace all particles in the uppermost and in the lowermost layer parallel to the center plane, as seen from the direction of magnetization. These outermost layers are the $n$-th layers of particles as counted from the $0$-th layer, the latter referring to the center plane. $n\in\mathbb{N}$ depends on the specific regular particle arrangement at hand. To further increase $x$, we instead replace the larger particles in each $(n-1)$-th, $(n-2)$-th, ..., third, second, and every layer by smaller particles, again counted from the center plane. 

\begin{figure}
\includegraphics[width=8.3cm]{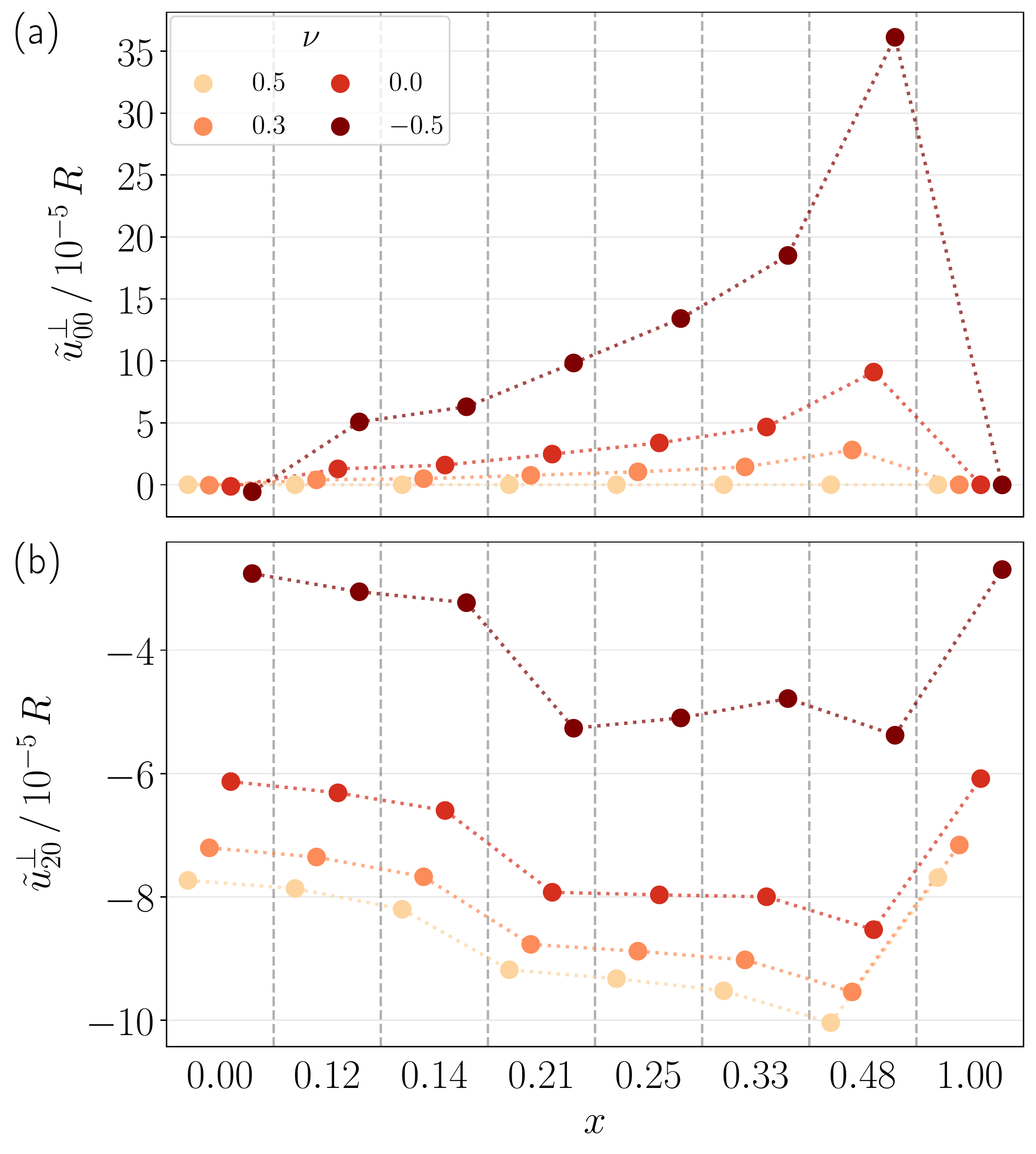}
\caption{Same as in figure~\ref{fig_sc_rand} for a simple cubic lattice structure, but here increasing $x$ through a layerwise replacement of larger by smaller particles. First, all particles in the center plane normal to the axis of magnetization are replaced. Then, additionally, every $n$-th, $(n-1)$-th, ..., third, second, and each layer of particles is replaced, where $n\in\mathbb{N}$ refers to the outermost layers as counted from the center plane. There are pronounced effects nonmonotonic with $x$ concerning the rescaled magnitudes of deformation. (a) A very large rescaled increase in total volume is observed for the evaluated Poisson ratios $\nu<0.5$ when every second layer of larger particles is replaced. (b) With increasing $x<1$, the rescaled magnitudes of contraction along the axis of magnetization are found to increase significantly and monotonically, except for $\nu=-0.5$.}
\label{fig_sc_layer}
\end{figure}

\begin{figure}
\includegraphics[width=8.3cm]{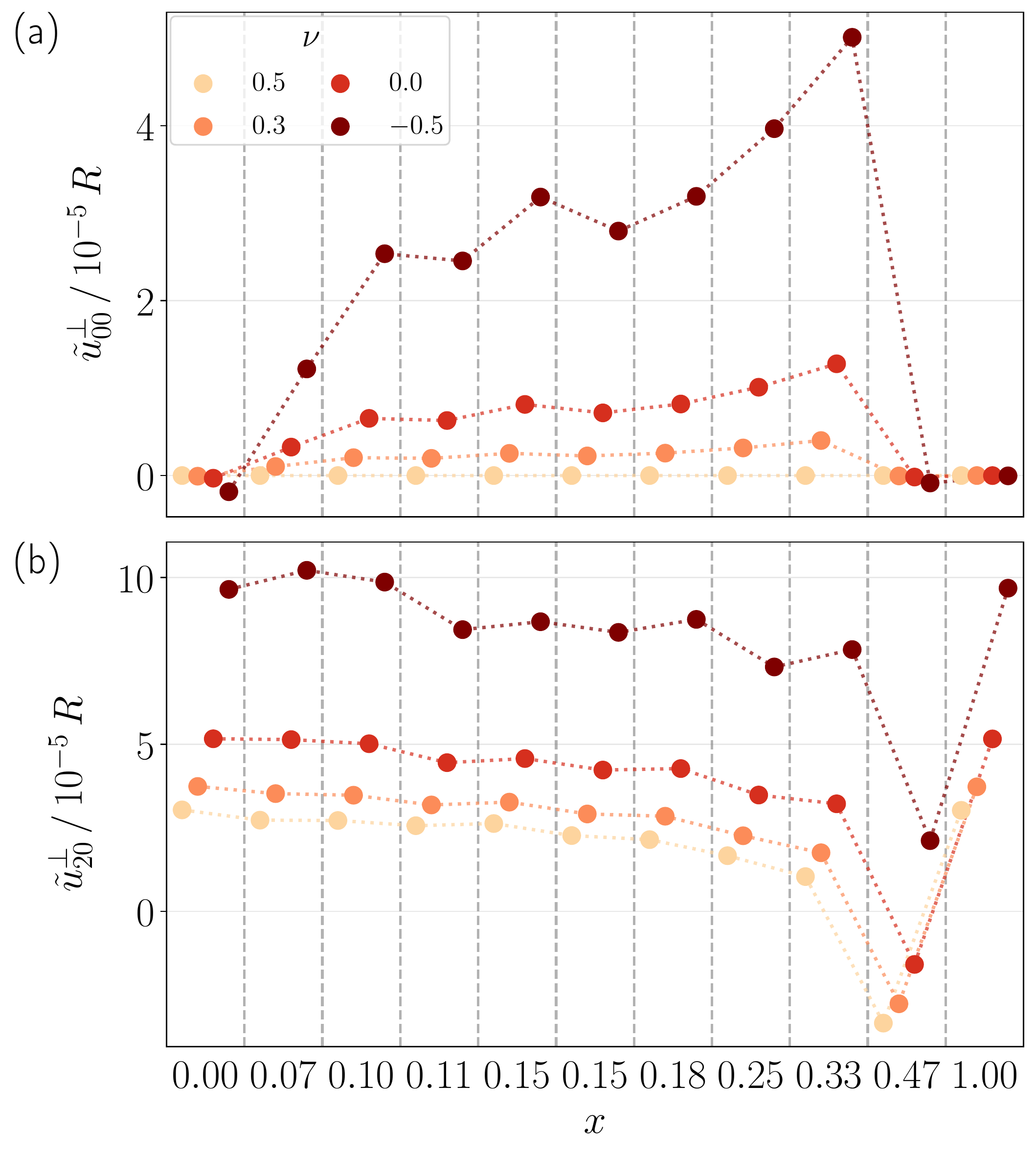}
\caption{Same as in figure~\ref{fig_sc_layer}, but for a body-centered cubic lattice structure. (a) Similar results are obtained as for the simple cubic structures, but here with the maximum of total increase in volume observed when all particles in every third layer are replaced. (b) The rescaled magnitude of elongation along the magnetization axis mainly decreases with increasing $x<1$, with slight nonmonotonicities. When every second layer of particles is replaced, even an inversion of the behavior into contraction along the magnetization axis is observed for the evaluated Poisson ratios $\nu>-0.5$ at $x\approx0.47$.}
\label{fig_bcc_layer}
\end{figure}

\begin{figure}
\includegraphics[width=8.3cm]{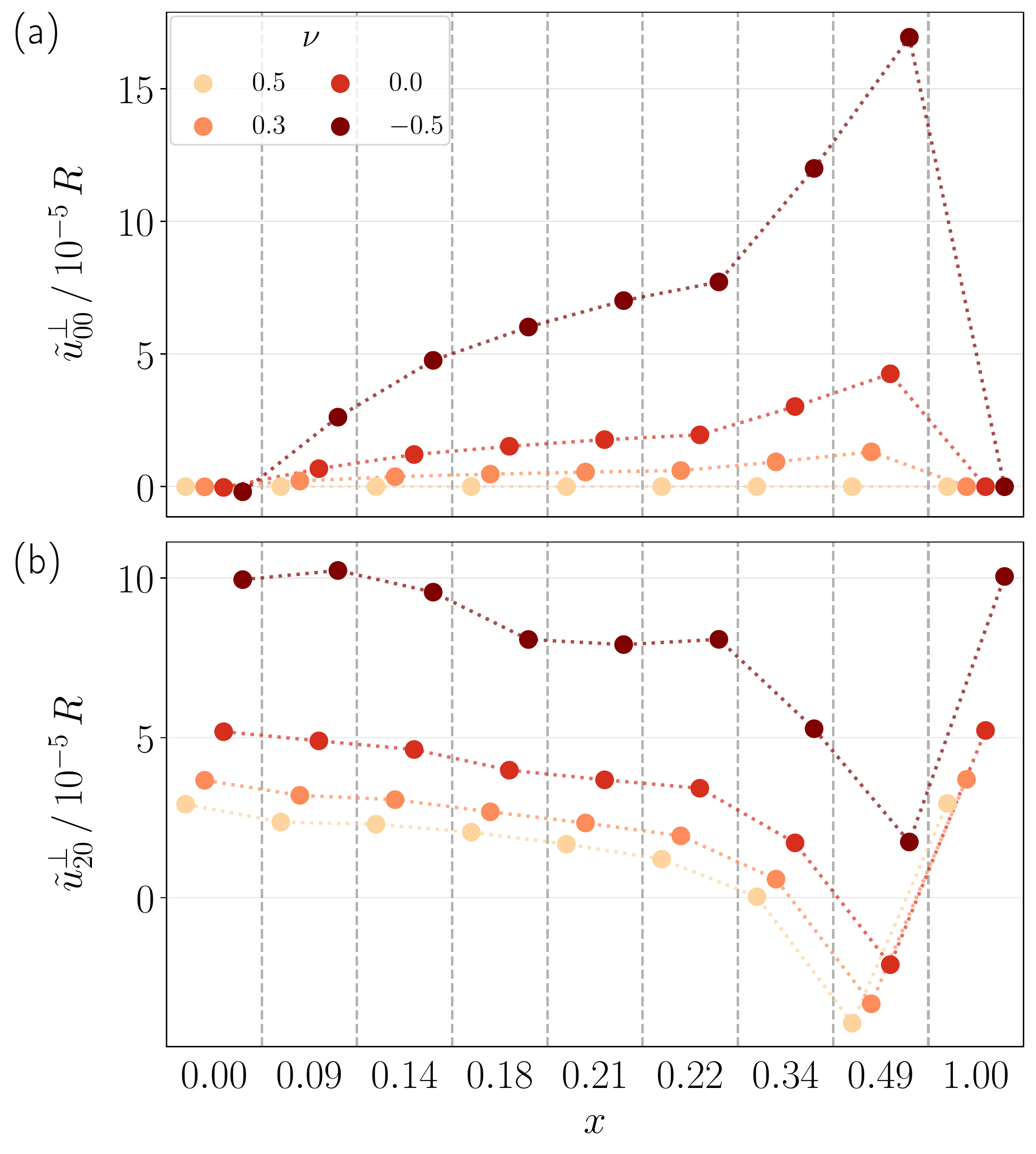}
\caption{Same as in figure~\ref{fig_sc_layer}, but for a face-centered cubic lattice structure. The results are very similar to those for the body-centered cubic lattice structure in figure~\ref{fig_bcc_layer}, only that the maximal rescaled increase in total volume in (a) is observed for the evaluated Poisson ratios $\nu>0.5$ when every second instead of every third layer of particles is replaced.}
\label{fig_fcc_layer}
\end{figure}

\begin{figure}
\includegraphics[width=8.3cm]{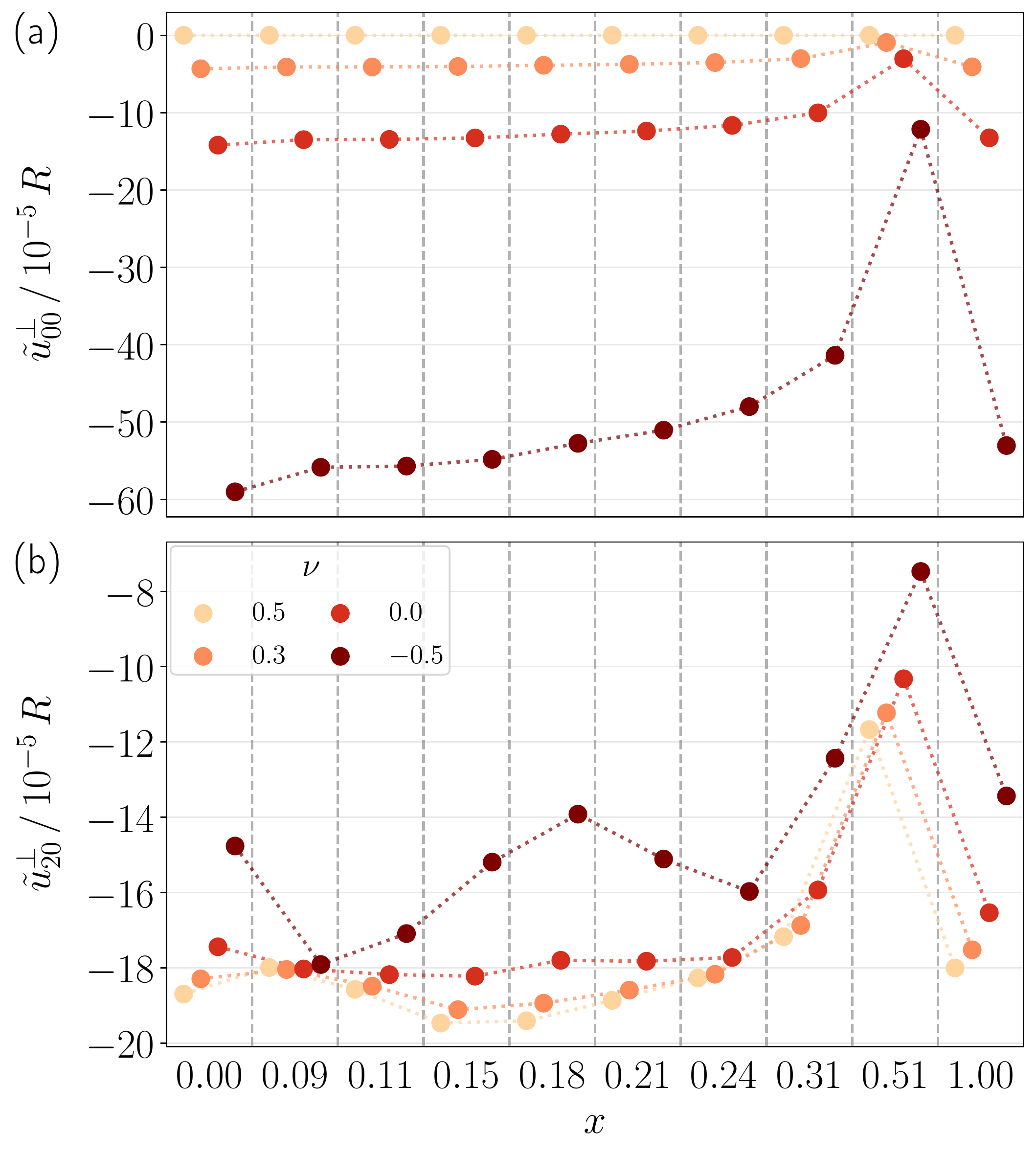}
\caption{Same as in figure~\ref{fig_sc_layer}, but for quadratically arranged chain-like aggregates aligned along the axis of magnetization. (a) Here, the layer-by-layer replacement of larger by smaller particles with increasing $x<1$ reduces the rescaled magnitude of the decrease in total volume for the evaluated Poisson ratios $\nu<0.5$. The effect is markedly pronounced when every second layer is replaced at $x\approx0.51$. (b) Similarly, the rescaled magnitude of overall relative contraction along the magnetization axis is severely reduced for $x\approx0.51$. The behavior is nonmonotonic as a function of $x$. Interestingly, larger rescaled magnitudes of overall relative contraction along the axis of magnetization than those at $x=0$ are observed at several values of $0<x<0.3$ for all evaluated Poisson ratios.}
\label{fig_quadchain_layer}
\end{figure}

\begin{figure}
\includegraphics[width=8.3cm]{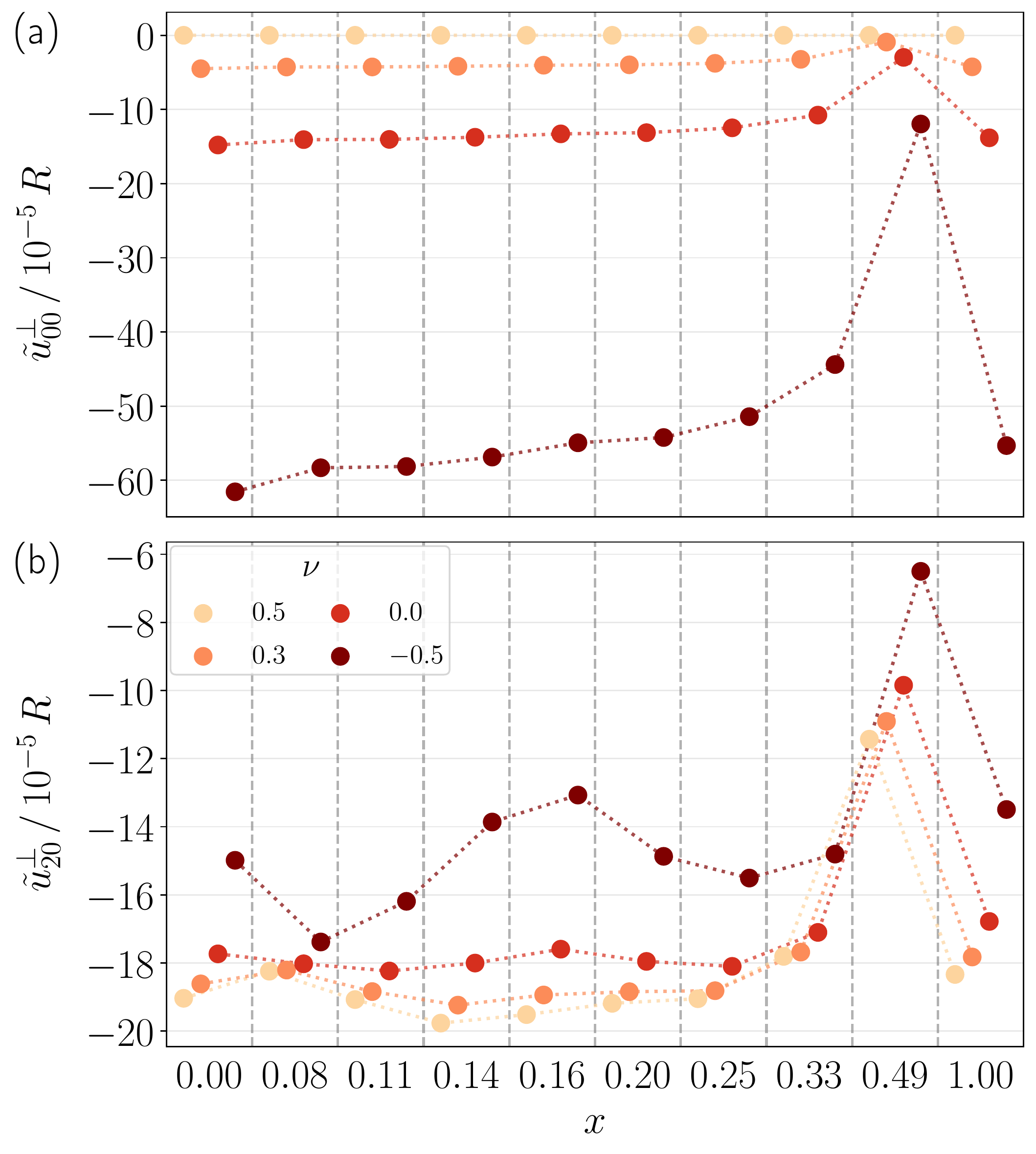}
\caption{Same as in figure~\ref{fig_sc_layer}, but for hexagonally arranged chain-like aggregates aligned along the axis of magnetization. The observed types of behavior are similar to the ones displayed for the quadratically arranged chain-like aggregates in figure~\ref{fig_quadchain_layer}.}
\label{fig_hexchain_layer}
\end{figure}

We studied the consequences of such layerwise replacement of larger by smaller particles for the simple cubic, body-centered cubic, and face-centered cubic lattice structures as well as for the quadratically and hexagonally arranged chain-like structures, see sections~\ref{sec_sc}, \ref{sec_bcc}, \ref{sec_fcc}, \ref{sec_quadchain}, and \ref{sec_hex_chain}, respectively. Corresponding results are depicted in figures~\ref{fig_sc_layer}--\ref{fig_hexchain_layer}. 

Concerning all regular cubic lattice arrangements, we find that the layerwise replacement of larger by smaller particles can significantly enhance the rescaled magnetically induced increase in total volume, see figures~\ref{fig_sc_layer}(a), \ref{fig_bcc_layer}(a), and \ref{fig_fcc_layer}(a). For the investigated Poisson ratios $\nu<0.5$, this effect is largest for the simple cubic and face-centered cubic lattices when every second layer is replaced. In contrast to that, for the body-centered cubic lattice it is most pronounced when every third layer is replaced. 

Moreover, for all these three lattice types the value of $\tilde{u}^{\bot}_{20}$ shows a nonmonotonic behavior as a function of $x$. We observe in each case a pronounced minimum when every second layer of larger particles is replaced by smaller ones, see figures~\ref{fig_sc_layer}(b), \ref{fig_bcc_layer}(b), and \ref{fig_fcc_layer}(b). For the system containing the simple cubic lattice structure this implies that the rescaled magnitude of relative contraction along the axis of magnetization is largest in this case. Conversely, for the body-centered and face-centered cubic structures this implies a reduced rescaled magnitude of relative extension along the axis of magnetization. When every second layer of particles is replaced, we even observe a reversed behavior for the evaluated Poisson ratios $\nu>-0.5$ for the body-centered and face-centered cubic structures. That is, these systems show a relative contraction along the axis of magnetization instead of relative extension, see figures~\ref{fig_bcc_layer}(b) and \ref{fig_fcc_layer}(b). 

The results for the quadratically and hexagonally arranged chain-like aggregates are relatively similar to each other, see figures~\ref{fig_quadchain_layer} and \ref{fig_hexchain_layer}. First, the magnetically induced rescaled reduction in overall volume significantly decreases in magnitude with increasing $x<1$ for the evaluated Poisson ratios $\nu<0.5$. A most pronounced reduction in magnitude is found when every second layer of larger particles is replaced by smaller ones. Second, the rescaled relative contraction along the axis of magnetization varies nonmonotonically with $x$. Also this effect is most severely reduced in magnitude when the replacement of particles occurs in every second layer. Yet, remarkably, we here again observe for all evaluated Poisson ratios that the effect is increased in magnitude above the one at $x=0$ for some intermediate values of $0<x<0.3$. This confirms that, indeed, the combination of different particle sizes can enhance the magnitude of magnetically induced contraction per squared employed mass of magnetizable material.

\section{Conclusions}
\label{sec_concl}

In summary, we have investigated the magnetically induced overall elastic deformation of spherical model systems of magnetic gels and elastomers containing discrete arrangements of magnetizable particles of binary size distribution. Simple cubic, body-centered cubic, face-centered cubic, and randomized isotropic particle arrangements were studied as well as systems containing quadratically or hexagonally arranged straight chain-like particle aggregates or globally twisted chain-like structures. In each case, we systematically increased the fraction of smaller particles at the cost of larger particles, keeping the total number of particles constant. Additionally, the role of the compressibility of the elastic matrix material was analyzed. We concentrate on the change of the overall volume in response to a saturating homogeneous external magnetic field as well as on the amount of relative extension or contraction along the axis of magnetization. For the systems containing the twisted chain-like structures we further evaluate the magnetically induced overall torsional deformation. To be able to compare the results for different number fractions of larger and smaller particles, we appropriately rescaled the magnitudes of the induced overall deformational response. 

Our results indicate that completely random replacements of larger by smaller particles in the investigated particle arrangements do not significantly affect the averaged overall rescaled deformational response. Only for the evaluated systems containing chain-like structures, slight trends of increased averaged rescaled mechanical reaction were observed as a consequence of randomly assigned binary particle sizes. However, specific individual realizations of systems of mixed particle sizes can show remarkably different types of behavior. Therefore, selectively replacing larger by smaller particles in a targeted approach allows to design the nature of the overall rescaled deformational response. Even qualitative changes are possible, for example, relative magnetically induced contractions along the axis of magnetization can be reversed into relative expansions, and vice versa. The effect is solely tuned by selectively positioning particles of different sizes onto the particle sites. Finally, these trends were confirmed when we studied the consequences of layerwise exchange of particle sizes, which may be important for subsequent steps of practical automated realizations of corresponding systems. 

Although we here were presenting our results in the context of magnetically induced deformations of magnetic gels and elastomers, our discussion equally applies to electrically induced deformations of electrorheological gels and elastomers when exposed to homogeneous external electric fields \cite{an2003actuating, allahyarov2015simulation, liu2001electrorheology}. In this case, the inclusions are electrically polarized by the external field and their mutual interactions are described by the formally identical electric dipolar interactions \cite{jackson1962classical}. The analogy holds as long as dynamic effects like electrically induced leakage currents do not play a significant role. 

Finally, we hope that our investigation will stimulate the further research into the controlled use of mixed particle sizes to optimize the overall material behavior \cite{li2010study}. Besides direct fabrication of samples containing randomized or uniaxially structured arrangements, also positioning into requested other discrete patterns might become possible in the future when synthesizing the materials. Then the selective positioning of particles of varying sizes on specific target locations may become an additional means to optimize and adjust the resulting desired overall material behavior.

\begin{acknowledgments}
The authors gratefully acknowledge support by the Deutsche Forschungsgemeinschaft through the priority program SPP 1681, grant no.~ME 3571/3. Some of the  results in this paper were derived using the HEALPix package \cite{HEALPix}.
\end{acknowledgments}


%

%

%
\end{document}